\documentclass[twocolumn]{aastex631}

\usepackage{threeparttable}
\usepackage{graphicx}	
\usepackage{amsmath}	
\usepackage{float}
\usepackage{comment}
\usepackage{xurl}

\newcommand{\um}{$\mu$m}
\newcommand{\water}{H$_2$O}
\newcommand{\methane}{CH$_4$}

\newcommand{\carbondiox}{CO$_2$}

\begin{document}

\title{A Pair of Dynamically Interacting Sub-Neptunes Around TOI-6054}

\author[0009-0002-2757-4138]{Maxwell A. Kroft}
\author[0000-0002-9539-4203]{Thomas G. Beatty}
\affil{Department of Astronomy, University of Wisconsin--Madison, 475 N. Charter Street, Madison, WI, 53706, USA}

\author{Ian J. M. Crossfield}
\affil{Department of Physics and Astronomy, University of Kansas, Lawrence, KS, USA}
\affil{Max-Planck-Institut f{\"u}r Astronomie, K{\"o}nigstuhl 17, D-69117 Heidelberg, Germany}

\author[0000-0003-3888-3753]{Joseph R. Livesey}
\author[0000-0002-7733-4522]{Juliette Becker}
\affil{Department of Astronomy, University of Wisconsin--Madison, 475 N. Charter Street, Madison, WI, 53706, USA}

\author[0000-0002-4927-9925]{Jacob K. Luhn}
\affil{Jet Propulsion Laboratory, California Institute of Technology, 4800 Oak Grove Drive, Pasadena, California 91109}

\author[0000-0003-0149-9678]{Paul Robertson}
\affil{Department of Physics and Astronomy, University of California, Irvine, CA 92697, USA}

\author[0000-0001-6637-5401]{Allyson Bieryla}
\affil{Center for Astrophysics \textbar \ Harvard \& Smithsonian, 60 Garden Street, Cambridge, MA 02138, USA}

\author[0000-0002-5741-3047]{David~R.~Ciardi}
\affil{NASA Exoplanet Science Institute-Caltech/IPAC, Pasadena, CA 91125, USA}

\author[0000-0002-2361-5812]{Catherine A. Clark}
\affil{NASA Exoplanet Science Institute-Caltech/IPAC, Pasadena, CA 91125, USA}

\author[0000-0003-2228-7914]{Maria V. Goliguzova}
\affil{Sternberg Astronomical Institute, Lomonosov Moscow State University, Moscow, Russia, 119992}

\author[0000-0002-2532-2853]{Steve~B.~Howell}
\affil{NASA Ames Research Center, Moffett Field, CA 94035, USA}

\author[0000-0001-6513-1659]{Jack J. Lissauer}
\affiliation{NASA Ames Research Center, Moffett Field, CA 94035, USA}

\author[0000-0001-7746-5795]{Colin Littlefield}
\affiliation{Bay Area Environmental Research Institute, Moffett Field, CA 94035, USA}
\affiliation{NASA Ames Research Center, Moffett Field, CA 94035, USA}

\author[0000-0003-2527-1598]{Michael~B.~Lund}
\affil{NASA Exoplanet Science Institute-Caltech/IPAC, Pasadena, CA 91125, USA}

\author[0000-0003-1713-3208]{Boris S. Safonov}
\affil{Sternberg Astronomical Institute, Lomonosov Moscow State University, Moscow, Russia, 119992}

\author[0000-0001-8898-8284]{Joseph M. Akana Murphy}
\altaffiliation{NSF Graduate Research Fellow}
\affil{Department of Astronomy and Astrophysics, University of California, Santa Cruz, CA 95064, USA}

\author{Natalie M. Batalha}
\affil{Department of Astronomy and Astrophysics, University of California, Santa Cruz, CA 95064, USA}

\author{Malik Bossett}
\affil{Department of Astronomy and Astrophysics, University of California, Santa Cruz, CA 95064, USA}

\author{Jonathan Brande}
\affil{Department of Physics and Astronomy, University of Kansas, Lawrence, KS, USA}

\author[0000-0002-6939-9211]{Tansu Daylan}
\affiliation{Department of Physics and McDonnell Center for the Space Sciences, Washington University, St. Louis, MO 63130, USA}

\author{Courtney Dressing}
\affil{Department of Astronomy, University of California Berkeley, Berkeley, CA 94720, USA}

\author{Anna Gagnebin}
\affil{Department of Astronomy and Astrophysics, University of California, Santa Cruz, CA 95060, USA}

\author{Daniel Huber}
\affil{Institute for Astronomy, University of Hawai’i, 2680 Woodlawn Drive, Honolulu, HI 96822, USA}
\affil{Sydney Institute for Astronomy
(SIfA), School of Physics, University of Sydney, NSW 2006, Australia}

\author[0000-0002-0531-1073]{Howard Isaacson}
\affil{Department of Astronomy, University of California Berkeley, Berkeley, CA 94720, USA}

\author[0000-0002-7084-0529]{Stephen R. Kane}
\affil{Department of Earth and Planetary Sciences, University of California, Riverside, CA 92521, USA}

\author{Laura Kreidberg}
\affil{Max-Planck-Institut f{\"u}r Astronomie, K{\"o}nigstuhl 17, D-69117 Heidelberg, Germany}

\author{David W. Latham}
\affil{Center for Astrophysics \textbar \ Harvard \& Smithsonian, 60 Garden Street, Cambridge, MA 02138, USA}

\author[0000-0002-4671-2957]{Rafael Luque}
\altaffiliation{NHFP Sagan Fellow}
\affil{Department of Astronomy \& Astrophysics, University of Chicago, Chicago, IL 60637, USA}

\author{Alex S. Polanski}
\affil{Department of Physics and Astronomy, University of Kansas, Lawrence, KS, USA}
\affil{Lowell Observatory, 1400 W. Mars Hill Rd., Flagstaff, AZ, 86001, USA}

\author[0000-0001-5728-4735]{Pranav H. Premnath}
\affil{Department of Physics and Astronomy, University of California, Irvine, CA 92697, USA}

\author{Maleah Rhem}
\affil{Department of Physics and Astronomy, University of Kansas, Lawrence, KS, USA}

\author[0009-0005-5520-1648]{Claire J. Rogers}
\affil{Department of Physics and Astronomy, University of California, Irvine, CA 92697, USA}

\author{Emma V. Turtelboom}
\affil{Department of Astronomy, University of California Berkeley, Berkeley, CA 94720, USA}

\begin{abstract}
We confirm the planetary nature of a pair of transiting sub-Neptune exoplanets orbiting the bright F-type sub-giant star TOI-6054 ($V=8.02$, $K=6.673$) as a part of the OrCAS radial velocity survey using WIYN/NEID observations. We find that TOI-6054b and TOI-6054c have radii of $2.65 \pm 0.15$ $R_{\oplus}$ and $2.81 \pm 0.18$ $R_{\oplus}$, respectively, and masses of $12.4 \pm 1.7$ $M_{\oplus}$ and $9.2 \pm 2.0$ $M_{\oplus}$. The planets have zero-albedo equilibrium temperatures of $1360 \pm 33$ K and $1144 \pm 28$ K. The host star has expanded and will evolve off of the Main Sequence within the next $\sim$500 Myr, and the resulting increase in stellar luminosity has more than doubled the stellar flux the two planets receive compared to the start of the host star's main sequence phase. Consequently, TOI-6054b may be losing some of its primordial H/He atmosphere -- if it has one. Based on dynamical simulations performed using the orbital parameters of the two planets, TOI-6054b, and TOI-6054c are very likely in a 5:3 mean motion resonance. The TOI-6054 system thus has the potential to be an excellent candidate for future atmospheric follow-up observations, with two similarly sized sub-Neptunes around a bright star. We also estimate that if TOI-6054b is currently losing its H/He atmosphere this should be observable from space and from the ground.
\end{abstract}

\section{Introduction} \label{sec:intro}

One of the first surprises of exoplanet demographics studies was the prevalence of planets larger than Earth and smaller than Neptune, many of which are on short period orbits, with Sun-like stars having an average of at least one such planet \citep{FP18}. Even more confounding was the discovery of a valley in the radius distribution of such planets spanning from 1.5 to 2.0 $R_{\oplus}$, and with peaks at 1.2 and 2.4 $R_{\oplus}$ \citep{FP18, Ve2018}. The smaller and larger groups of these intermediate sized planets are generally known now as super-Earths and sub-Neptunes, respectively.

This radius valley is thought to be sculpted by atmospheric mass loss during the early stages of planetary evolution \citep{Owen_Wu_2013}. This atmospheric loss produces a dip in the radius distribution of small planets between those that have been completely stripped (super-Earths) and those which are massive enough to have retained an envelope (sub-Neptunes) \citep{Owen_Wu_2017}. The two leading theories to explain this radius valley are atmospheric photoevaporation \citep[e.g.,][]{Owen_Wu_2017} and core-powered mass-loss \citep[e.g.,][]{core_mass_loss}, and recent work suggests that both are responsible in varying regimes of parameter space \citep{Owen_Schlichting_2024}. Other works suggest that the radius valley can be produced naturally through primordial gas accretion alone, even in the absence of mass loss, while simultaneously reproducing the observed population of large planets \citep{Lee_Karalis_Thorngren_2022}.

Super-Earths and sub-Neptunes are thought to be separated not only in radius, but also in bulk composition. Super-Earths are consistent with a primarily rock and iron core composition and relatively thin atmospheres \citep{Zeng19}. However, sub-Neptune compositions remain poorly constrained, as their bulk densities are consistent with both water worlds and rocky cores with H/He dominated atmospheres \citep{Rogers23}. It is also possible that many of these sub-Neptunes have liquid water oceans and an H/He dominated atmosphere, referred to as Hycean worlds \citep{hycean}.

Sub-Neptunes, especially those on short period orbits, are unlike any solar system planets, and thus are of particular interest to planet formation studies. Some of their potential compositions indicate formation beyond the water ice line, and likely require type I migration to reach close orbits \citep{Venturini20}. Additionally, \cite{Weiss2018} has found that the primary formation outcome of multi-planet systems, most of which consist of super-Earths and sub-Neptunes, is a set of similarly sized planets with regular orbital spacing. Sub-Neptunes show evidence of complex formation histories, and atmospheric observations provide the best avenue towards unveiling these formation histories as well as breaking the compositional degeneracy \citep[e.g.,][]{Deming_Seager_2017}.

Currently, only five sub-Neptune-sized planets have atmospheric constraints. K2-18b, a possible Hycean world, has robust measurements of methane and carbon dioxide in a hydrogen rich atmosphere \citep{k218b}. \cite{toi270d} observed strong signals of methane, carbon dioxide, and water in the atmosphere of TOI-270d, a sub-Neptune in a multi-planet system with two inner transitting planets, and infer that it has a primarily rock/iron core. GJ 3470 b is a Neptune-sized planet with evidence of disequilibrium photochemistry and significant detections of water, methane, sulfur dioxide, and carbon dioxide \citep{gj3470}. GJ 9827 d is thought to be a water world, with a highly significant detection of water and an inferred volume mixing ratio of water of greater than 31\% \citep{gj9827d}. GJ 1214 b has moderate evidence carbon dioxide and methane, and multiple works support an atmospheric metallicity greater than 1000x solar \citep{GJ1214_Gao,GJ1214_Kempton,GJ1214_Schlawin}. Additionally, atmospheric non-detections by the JWST COMPASS program have ruled out low metallicity, cloudless atmospheres in TOI-836b \citep{836b}, TOI-836c \citep{836c}, L 98-59 c \citep{L98-59c}, and L168-9 b \citep{L168-9b}. So far, the community has already observed a great diversity in the atmospheric compositions of sub-Neptunes, further supporting the varied formation and evolution pathways hypothesized to produce this class of planet. However, in order to continue to increase this sample, more targets with well measured masses are required.

In this paper we present the discovery and characterization of an exciting pair of transiting sub-Neptune exoplanets, orbiting the sub-giant star TOI-6054. These planets provide a rare opportunity to study the bulk compositions and evolutionary history of two sub-Neptunes which formed in the same environment.

\section{Observations} \label{sec:obs}

TOI-6054 was selected for follow-up observations as part of the OrCAS survey, a radial velocity survey to inform our understanding of the Origins, Compositions, and Atmospheres of sub-Neptune exoplanets, defined in \cite{orcas}. This program aims to provide precise masses for a large number of sub-Neptunes, in order to facilitate atmospheric characterization of a demographically representative sample of these planets with JWST \citep{jwst} and soon, ARIEL \citep{ariel}.

\subsection{TESS Photometry} \label{sec:tess}

The Transiting Exoplanet Survey Satellite (TESS, \citealt{TESS}) observed the star TIC 392681545 as part of its all-sky survey in Sectors 19 and 59. TESS performed the Sector 19 observations at a 30 minute cadence and these observations spanned from UT 2019 Nov 27 to UT 2019 Dec 24. TESS performed the Sector 59 observations at a 2 minute cadence and these observations spanned from UT 2022 Nov 26 to UT 2022 Dec 23. The TESS team performed data reduction on the observations with the TESS Science Processing Operations Center (SPOC, \citealt{SPOC}) pipeline. They detected potential transits and promoted the system to a TESS object of interest (TOI) named TOI-6054. We use SPOC data products resampled to a 200 second cadence for Sector 59.

The Sector 19 photometry shows some variability, likely due to changing stellar brightness or TESS systematics, while the Sector 59 photometry is relatively stable (see Figure \ref{fig:det}, top panels). We removed this variability using damped harmonic oscillator Gaussian Process kernels during fitting, which is discussed further in Section \ref{sec:orbpar}. Additionally, we removed 94 of 1203 and 1,767 of 11,417 points from the Sector 19 and Sector 59 data, respectively, due to any bad quality flags and nan values in the flux data \citep{spoc_dat_products}. TOI-6054b has four transits in each sector of depth $\sim$215 ppm, and TOI-6054c has two transits in each sector of depth $\sim$245 ppm.

\subsection{NEID Spectroscopy} \label{sec:neid}

We obtained twenty-nine radial velocity (RV) measurements of the TOI-6054 system with the NEID spectrograph \citep{NEID}. NEID is a high resolution (R$\sim$115,000) spectrograph on the WIYN 3.5m telescope at Kitt Peak, Arizona. We performed our observations between UT 2023 Nov 14 and UT 2024 November 18. We used the standard NEID data reduction pipeline version 1.4.0\footnote{\url{https://neid.ipac.caltech.edu/docs/NEID-DRP/}} as implemented by the NEID Archive to extract velocity measurements from these observations. We took the first four observations using an 800 second exposure time, and the remaining twenty-five using a 1200 second exposure time once we determined that there was a measurable RV signal in the data. The RV measurements had an average precision of 1.6 m/s and a standard deviation of 3.1 m/s, and are listed in Table \ref{tab:rv}.

\begin{table}[]
\centering
\caption{Radial velocity measurements of TOI-6054 used in this paper. We have added 14,733.25 m/s to all RV measurements listed here and used in this work, to center the data near 0 m/s.}
    \begin{tabular}{ll}
        \hline
        Time (BJD-2450000) & RV (m/s) \\
        \hline
        10262.70438056 & $-6.65 \pm 3.3$ \\ 
        10270.68230503 & $1.35 \pm 2.2$ \\ 
        10282.66410366 & $-3.85 \pm 1.6$ \\ 
        10288.8804618 & $-3.15 \pm 1.5$ \\ 
        10294.60807495 & $0.65 \pm 1.7$ \\ 
        10299.59925936 & $-3.05 \pm 1.4$ \\ 
        10300.67144573 & $-1.65 \pm 1.7$ \\ 
        10307.61006737 & $1.95 \pm 1.3$ \\ 
        10310.64678688 & $-2.35 \pm 1.4$ \\ 
        10318.61108143 & $-0.45 \pm 2.0$ \\ 
        10319.62536317 & $-2.35 \pm 1.1$ \\ 
        10320.6349887 & $-5.45 \pm 1.4$ \\ 
        10322.81399018 & $-1.75 \pm 1.6$ \\ 
        10323.63008892 & $-0.65 \pm 1.2$ \\ 
        10327.61693279 & $-1.55 \pm 1.2$ \\ 
        10328.64107165 & $-0.95 \pm 1.4$ \\ 
        10337.65005097 & $-0.25 \pm 2.0$ \\ 
        10338.66765365 & $1.95 \pm 1.7$ \\ 
        10386.60924854 & $-7.35 \pm 2.1$ \\ 
        10392.62903786 & $1.65 \pm 2.2$ \\ 
        10399.62698622 & $-2.55 \pm 2.6$ \\ 
        10585.01366493 & $-4.05 \pm 1.0$ \\ 
        10588.94727606 & $-4.65 \pm 0.9$ \\ 
        10590.74262762 & $-3.35 \pm 1.1$ \\ 
        10601.76825432 & $3.85 \pm 1.6$ \\ 
        10609.00148521 & $7.35 \pm 1.0$ \\ 
        10629.90177377 & $1.65 \pm 1.3$ \\ 
        10630.96209527 & $3.45 \pm 2.0$ \\ 
        10632.97205424 & $1.25 \pm 1.4$ \\ 
        \hline
    \end{tabular}
    \label{tab:rv}
    \begin{tablenotes}
        \item
    \end{tablenotes}
\end{table}

\subsection{High Resolution Imaging} \label{sec:img}

Spatially close stellar companions (bound or line of sight) can confound exoplanet discoveries such that the detected transit signal might be a false positive due to a background eclipsing binary. Even for real planet discoveries, the close companion will yield incorrect stellar and exoplanet parameters if unaccounted for \citep{ciardi2015,FurlanHowell2017,FurlanHowell2020}. Additionally, the presence of a close companion star may lead to the non-detection of small planets residing within the same exoplanetary system \citep{Lester2021}. In order to rule out these possibilities, we observed the TOI-6054 system with optical speckle observations at the Sternberg Astronomical Institute (SAI) and Gemini and near-infrared adaptive optics (AO) imaging at Palomar Observatory.

\begin{figure}
    \centering
    \includegraphics[width=\columnwidth]{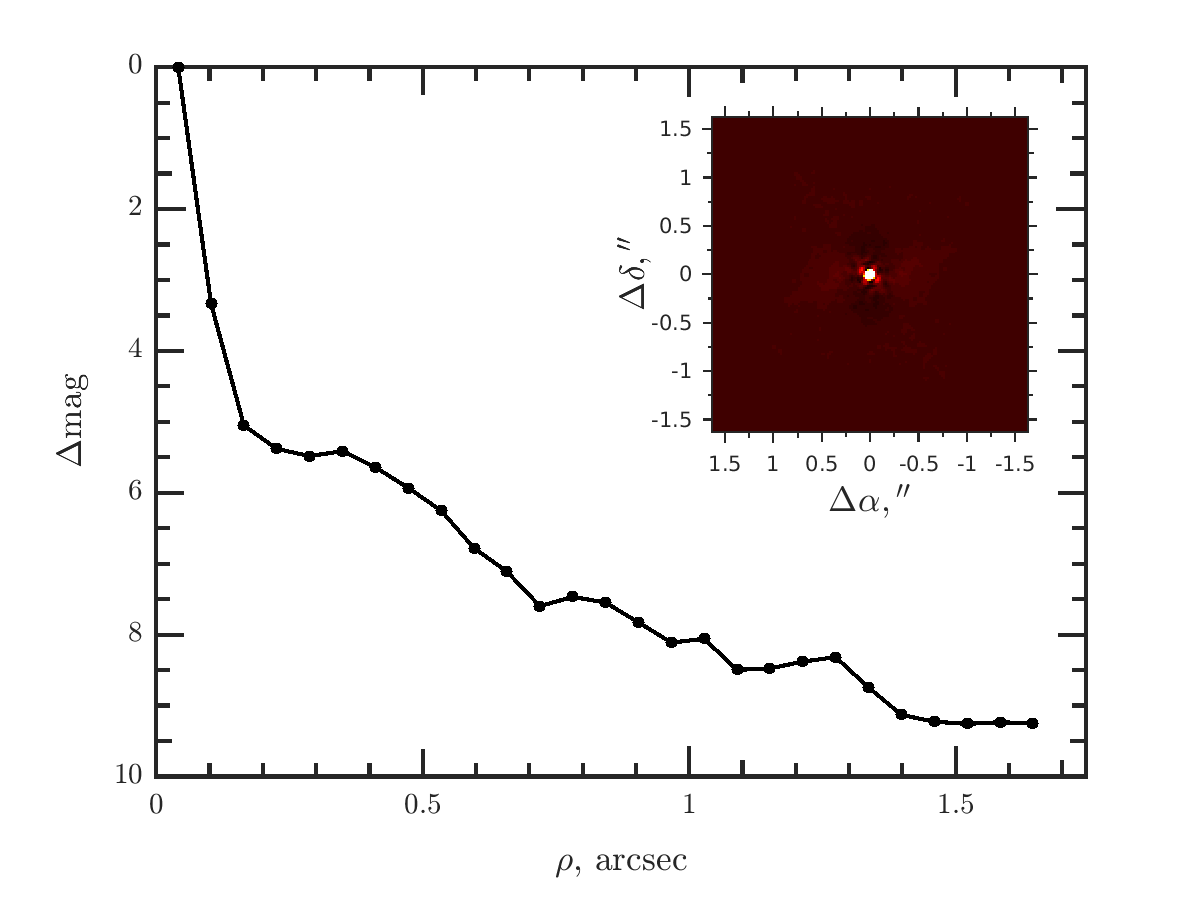}
    \includegraphics[width=0.9\columnwidth]{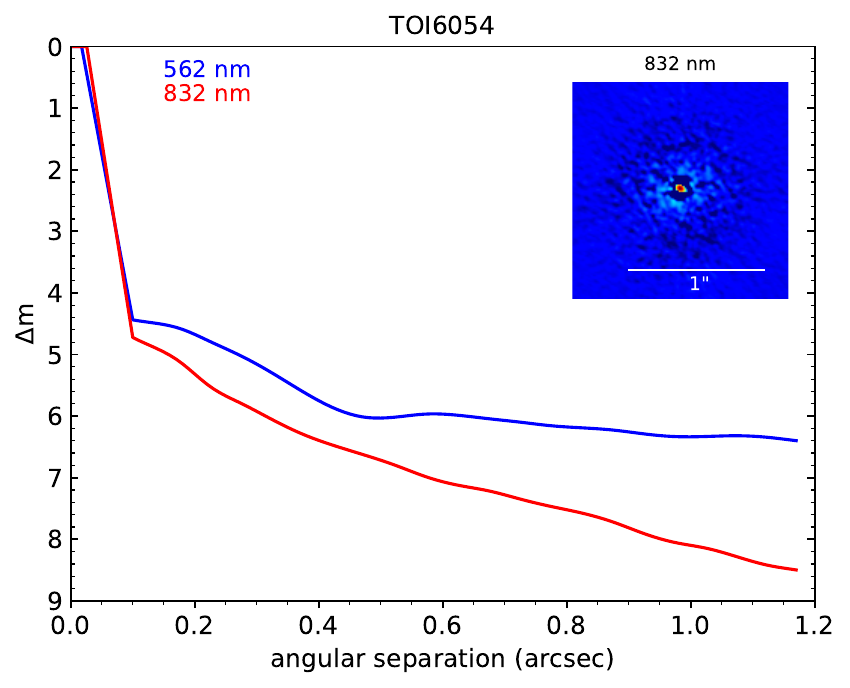}
    \includegraphics[width =\columnwidth]{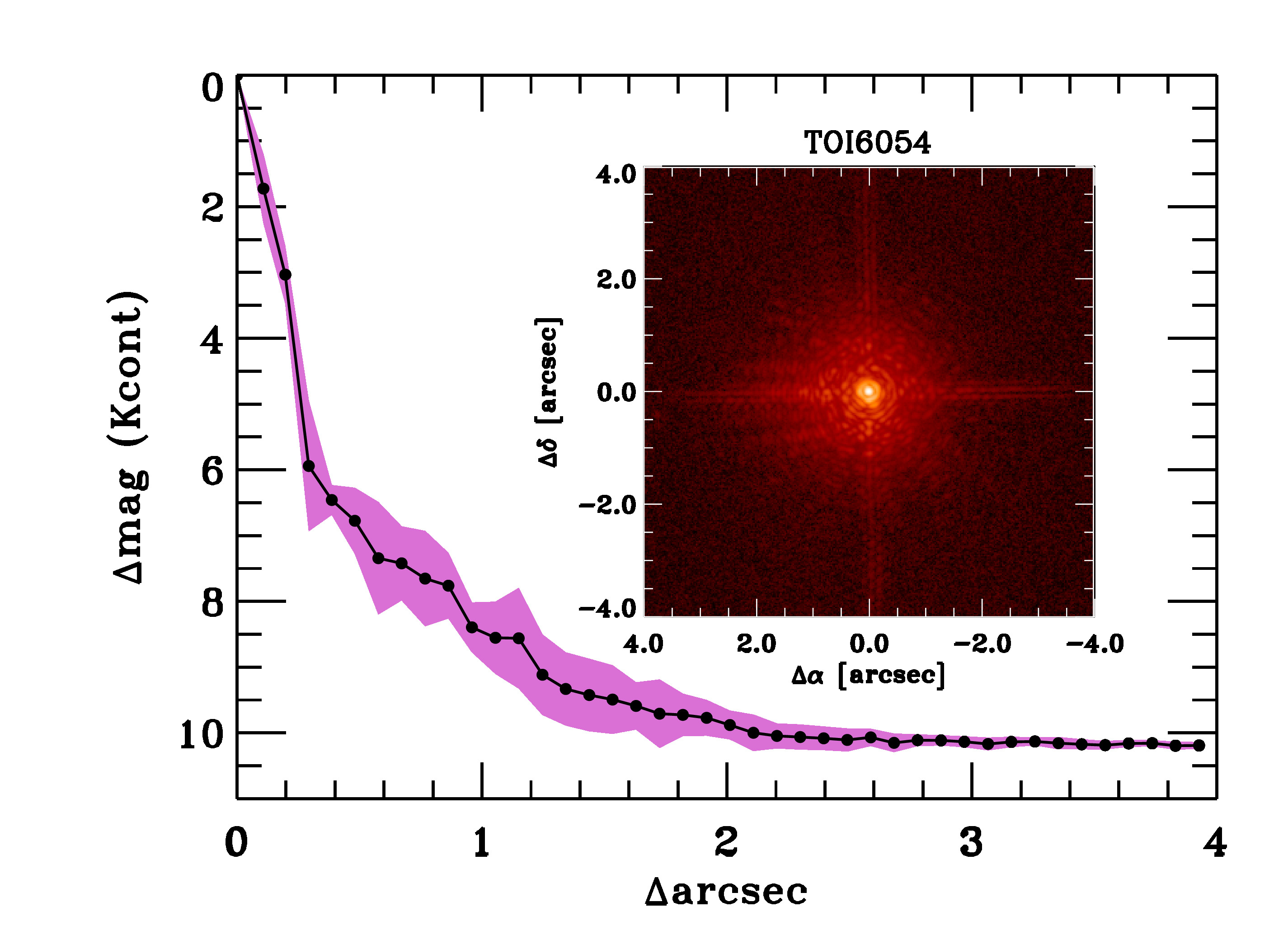}
    \caption{{\it Top}: Optical speckle imaging and sensitivity curves for the SAI observations, with a medium band filter centered on 625 nm. {\it Center}: Optical speckle imaging and sensitivity curves for the Gemini observations. The 562 nm sensitivity is shown in blue, and the 832 nm sensitivity curve is shown in red. {\it Bottom}: NIR AO imaging and sensitivity curves for the Palomar observations, in a narrowband Kcont filter centered on 2.29 $\mu$m. {\it Insets}: The central portions of each image.}
    \label{fig:image}
\end{figure}
	
\subsubsection{Optical Speckle Imaging}

We observed TOI-6054 with the SPeckle Polarimeter (SPP; \citealt{Strakhov23}) on UT 2023 Aug 27. SPP is a facility instrument on the 2.5m telescope at the Caucasian Observatory of SAI, Lomonosov Moscow State University. We used the fast, low-noise, CMOS Hamamatsu ORCA-quest detector with a medium band filter centered on 625 nm with a 50 nm FWHM. The pixel scale was 20.6 mas/pixel, the angular resolution was 63 mas, and the field of view was 5''x5'' centered on the star. We used the atmospheric dispersion compensator. The long-exposure seeing during acquisition was 0.64''. We estimated the power spectrum from 21,562 frames with 5.57 ms per exposure. We did not detect any stellar companions, with a detection limit of 5.4 mags and 8.0 mags at stellocentric distances of 0.25'' and 1.0'', respectively. At the distance of TOI-6054, these correspond to projected separations of 19.68 and 78.73 AU. The sensitivity curve is shown in the top panel of Figure \ref{fig:image}.

Additionally, we observed TOI-6054 on UT 2023 Dec 04 using the ‘Alopeke speckle instrument on the Gemini North 8m telescope \citep{Scott2021}.  ‘Alopeke provides simultaneous speckle imaging in two bands (562nm and 832 nm) with output data products including a reconstructed image with robust contrast limits on companion detections. Three sets of 1000-0.06 second images were obtained and processed using the standard reduction pipeline \citep{Howell2011}. The middle panel of Figure \ref{fig:image} shows the final contrast curves and the 832 nm reconstructed speckle image. We found that to within our achieved magnitude contrast TOI-6054 is a single star with no companion brighter than 5-8.5 magnitudes below that of the target star from the 8m telescope diffraction limit (20 mas) out to 1.2''. At the distance of TOI-6054 these angular limits correspond to projected separations of 1.6 to 95 AU.

\begin{figure*}
    \centering
    \includegraphics[width = \textwidth]{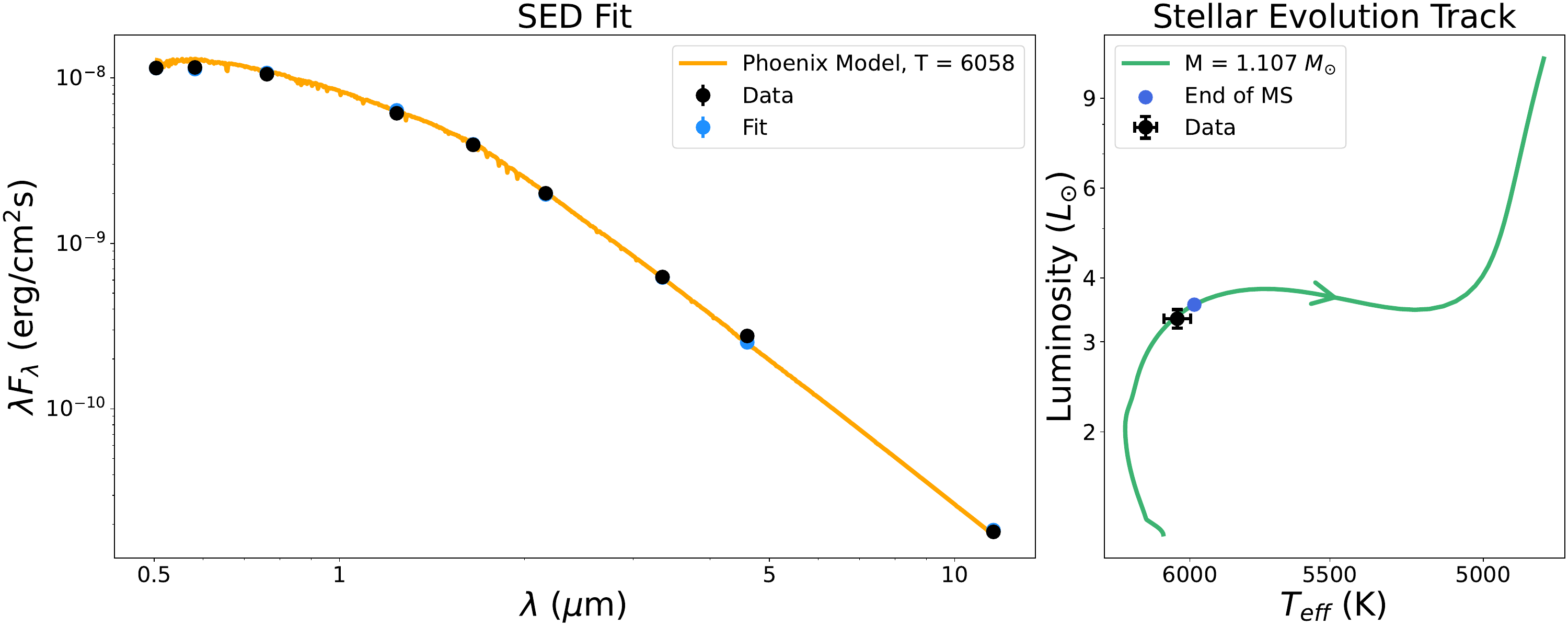}
    \caption{{\it Left}: The best fit stellar SED for TOI-6054. The data are in black, the best fits to the data are in blue, and the best fit stellar model is in orange. Error bars on both the data and the fit points are smaller than the points. {\it Right}: The stellar evolution track for the best fit stellar mass is plotted in green, with TOI-6054's $T_{eff}$ and $L_{bol,S}$ plotted in black. The end of the main sequence on this track, as defined by the MIST evolutionary phase, is plotted in blue. An arrow is plotted on the track indicating the forward direction in time.}
    \label{fig:sed}
\end{figure*}

\subsubsection{Near-Infrared AO Imaging}

We made observations of TOI-6054 on UT 2023 Nov 29 with the PHARO instrument \citep{hayward2001} on the Palomar Hale (5m) behind the P3K natural guide star AO system \citep{dekany2013} in the narrowband Kcont filter $(\lambda_o = 2.29\:\mu$m$; \Delta\lambda = 0.035\:\mu$m). The PHARO pixel scale is $0.025\arcsec$ per pixel. We performed a standard 5-point quincunx dither pattern with steps of 5\arcsec\ three times with each repeat separated by 0.5\arcsec. The reduced science frames were combined into a single mosaiced image with a final resolution of 0.096\arcsec.  The sensitivity of the final combined AO images were determined by injecting simulated sources azimuthally around the primary target every $20^\circ $ at separations of integer multiples of the central source's FWHM \citep{furlan2017}. The brightness of each injected source was scaled until standard aperture photometry detected it with $5\sigma $ significance.  The final $5\sigma $ limit at each separation was determined from the average of all of the determined limits at that separation and the uncertainty on the limit was set by the rms dispersion of the azimuthal slices at a given radial distance.  The Palomar sensitivities are shown in the bottom panel of Figure \ref{fig:image}. We found that TOI-6054 has no companion brighter than 8.4 magnitudes below that of the target star out to a distance of 1.0".

\subsection{Spectroscopic Parameters} \label{sec:sppar}

Two reconnaissance spectra were obtained on UT 2023 February 10 and 2023 March 9 using the Tillinghast Reflector Echelle Spectrograph (TRES, \citealt{gaborthesis}) mounted on the 1.5m Tillinghast Reflector telescope at the Fred Lawrence Whipple Observatory (FLWO) atop Mount Hopkins, Arizona. TRES is a fiber-fed echelle spectrograph with a wavelength range of 390-910nm and a resolving power of R~44,000. The spectra were extracted as described in \cite{buchhave2010} and then used to derive stellar parameters using the Stellar Parameter Classification tool (SPC, \citealt{buchhave2012}). SPC cross correlates an observed spectrum against a grid of synthetic spectra based on Kurucz atmospheric models \citep{kurucz1992} to derive effective temperature $T_{eff}$, surface gravity log $g_{s}$, metallicity [Fe/H], and rotational velocity $v$sin $i_{s}$ of the star. We report the derived stellar parameters from the 2023 March 9 spectra in Table \ref{tab:stpar}, as this observation had a slightly high signal to noise ratio.

\newpage
\section{Fitting and Results} \label{sec:fit}

\subsection{SED Fitting} \label{sec:sed}

We fit the stellar SED of TOI-6054 to refine the estimated stellar radius, and hence our measurements of the planetary radii. We obtained photometric data for the spectral energy distribution (SED) of TOI-6054. We used the Gaia DR3 G\textsubscript{BP}, G, and G\textsubscript{RP} fluxes \citep{GaiaMission, GaiaDR3} as well as the Two Micron All Sky Survey (2MASS) J, H, and K\textsubscript{s} fluxes \citep{2MASS} and the Wide-Field Infrared Survey Explorer (WISE) WISE1, WISE2, and WISE3 fluxes \citep{WISE} reported in the Guide Star Catalog II \citep{GSC2}. We report these flux values in Table \ref{tab:stpar}.

\begin{table*}[]
\centering
\caption{Measured and derived stellar parameters for TOI-6054. Fit values reported are the median and the 68\% confidence interval.}
    \begin{tabular}{lllll}
        \hline
        Stellar Parameters \\
        \hline
        Identifiers & TOI-6054 \\
         & TIC 392681545 \\
         & HD 23074 \\
        2MASS & J03452100+6021033 \\
        Gaia DR3 & 474143193821024256 \\
        \hline
        Photometry$^{1}$ & Band & Flux (Jy)$^{2,3}$ & Magnitude$^{2,4,5}$\\
        & TESS & ... & 7.4961 $\pm$ 0.006\\
        & Gaia G\textsubscript{BP} & 1.93 $\pm$ 0.04 & 8.1643 $\pm$ 0.0028\\
        & Gaia G & 2.25 $\pm$ 0.05 & 7.8912 $\pm$ 0.0028\\
        & Gaia G\textsubscript{RP} & 2.68 $\pm$ 0.05 & 7.4498 $\pm$ 0.0038\\
        & 2MASS J & 2.53 $\pm$ 0.06 & 6.988 $\pm$ 0.026\\
        & 2MASS H & 2.17 $\pm$ 0.06 & 6.713 $\pm$ 0.029\\
        & 2MASS K\textsubscript{S} & 1.45 $\pm$ 0.03 & 6.673 $\pm$ 0.024\\
        & WISE1 & 0.700 $\pm$ 0.038 & 6.604 $\pm$ 0.059\\
        & WISE2 & 0.423 $\pm$ 0.012 & 6.515 $\pm$ 0.030\\
        & WISE3 & 0.0695 $\pm$ 0.0014 & 6.551 $\pm$ 0.020\\
        \hline
        Gaia Parameters$^{2}$ & RA (degrees) & Dec (degrees) & Parallax (mas) & $d$ (pc) \\
         & 56.3369915 & 60.3509164 & 12.701 $\pm$ 0.019 & 78.73 $\pm$ 0.12 \\
        \hline
        TrES Parameters & $T_{eff}$ (K) & log $g_{s}$ (cgs) & $v$sin $i_{s}$ (km/s) & [Fe/H] \\
         & 6047 $\pm$ 50 & 4.1 $\pm$ 0.1 & 6.69 $\pm$ 0.50 & -0.07 $\pm$ 0.08 \\
        \hline
        Derived Parameters \\
        \hline
        SED Fit & $R_{s}$ ($R_{\odot}$) & $A_{V}$ (mag) & $L_{bol,S}$ ($L_{\odot}$) \\ [5 pt]
        & $1.662_{-0.071}^{+0.071}$ & $0.038_{-0.025}^{+0.031}$ & $3.33_{-0.14}^{+0.14}$ \\ [5 pt]
        \hline
        MIST Isochrones$^{6}$ & $M_{s}$ ($M_{\odot}$) & $\rho_{s}$ (g/cm$^{3}$) & Age (Gyr)\\ [5 pt]
        & $1.107_{-0.040}^{+0.040}$ & $0.340_{-0.045}^{+0.045}$ & $6.0_{-1.1}^{+1.1}$ \\ [5 pt]
        \hline
    \end{tabular}
    \label{tab:stpar}
    \begin{tablenotes}
        \item $^{1}$ \cite{Vizier}. $^{2}$ \cite{GaiaMission,GaiaDR3}. $^{3}$ \cite{GSC2}. $^{4}$ \cite{vizier_2mass}. $^{5}$ \cite{vizier_wise}. $^{6}$ \cite{isochrones}.
    \end{tablenotes}
\end{table*}

We fit the stellar SED using Phoenix synthetic spectra \citep{Phoenix} assuming log $g_{s}$ = 4.0 and Solar metallicity. These values are consistent with the TrES measurements, and as demonstrated in \cite{Stevens18}, the precise values of log $g_{s}$ and metallicity assumed for this kind of broadband SED fitting have a minimal effect on the results. Since the Phoenix models step through Teff in 100 K intervals, we linearly interpolated between the synthetic points in each band. We corrected for extinction using the \cite{GCC09} extinction curve for the Gaia fluxes and the \cite{G21} extinction curve for the WISE data, both assuming R(V) = 3.1. We use the \texttt{dust\_extinction} package \citep{dust_extinction}, which calculates a fractional flux extinction with the aforementioned extinction curves given a wavelength and $A_{V}$, to convert our fitted $A_{V}$ into each band.

We used these extincted and interpolated Phoenix model fluxes to fit the broadband SED from TOI-6054. We performed the fit with four parameters: stellar effective temperature $T_{eff}$, distance $d$, stellar radius $R_{s}$, and visual band extinction $A_{V}$. We used Gaussian priors for $T_{eff}$ and $d$ according to the values from TrES and Gaia reported in Table \ref{tab:stpar}, and we recovered these priors in our posterior distributions.

We carried out the fitting process using the \texttt{emcee} package \citep{emcee}. We used ten walkers, with 2000 burn-in steps and 4000 production steps. To test for convergence, we checked that the autocorrelation times for each chain were less than 80 (the number of production steps divided by 50). We report the results of the SED fit in Table \ref{tab:stpar}, as well as $L_{bol,S}$, which we computed from $R_{s}$ and $T_{eff}$. We plot the best fit SED in the left panel of Figure \ref{fig:sed}.

In \cite{stellar_unc}, the authors show that uncertainties in interferometric angular diameters and bolometric fluxes set systematic uncertainty floors on stellar radii and luminosities. Following their recommendations, we increased the errors on our measured values of $R_{s}$ and $L_{bol,S}$ by adding a fractional error floor of 4.2\% and 2.4\%, respectively, in quadrature to the formal errors on $R_{s}$ and $L_{bol,S}$.

Based on this SED fitting, we estimate that TOI-6054 has a radius of $1.662 \pm 0.071$ $R_{\odot}$ and a bolometric luminosity of $3.33 \pm 0.14$ $L_{\odot}$.

\subsection{Stellar Isochrone Fitting} \label{sec:iso}

In order to measure accurate planetary masses, we also wished to accurately measure TOI-6054's mass. To do this, we interpolated our measured stellar parameters onto the MESA Isochrones and Stellar Tracks (MIST, \citealt{MIST1, MIST2, MESA1, MESA2, MESA3}) models. We used the \texttt{isochrones} python package \citep{isochrones}, which interpolates stellar models to photometric and spectroscopic data.

We fit to the MIST models using $T_{eff}$, log $g_{s}$, and [Fe/H] from the TrES spectroscopic parameters as well as the Gaia DR3 parallax. We also used photometric magnitudes in the same bands that we used for SED fitting (see Section \ref{sec:sed}, Table \ref{tab:stpar}). The resulting SED fits were consistent with our SED fit, and we adopt the measurements discussed in Section \ref{sec:sed} for the remainder of the analysis.

We then fit the MIST models using \texttt{MultiNest} \citep{multinest1,multinest2,multinest3}. In Table \ref{tab:stpar}, we report the median and 68\% confidence intervals for stellar mass $M_{s}$, average stellar density $\rho_{s}$, and stellar age.

\cite{stellar_unc} shows that uncertainties in stellar masses and ages are limited by uncertainties in stellar evolution models. Therefore, we increase the uncertainties on $M_{s}$ and stellar age following their recommendations. We interpolate over the MIST, YREC \citep{YREC}, DSEP \citep{DSEP}, and GARSTEC \citep{GARSTEC} stellar model grids given an input of $T_{eff}$ from TrES, $L_{bol,S}$ from the SED fit, and log $g_{s}$ from the \texttt{isochrones} fit using the package \texttt{kiauhoku} \citep{kiauhoku}. As suggested in \cite{stellar_unc}, we take the maximal difference between model grids as a systematic uncertainty floor of $\pm$0.028 $M_{\odot}$ in mass and $\pm$0.86 Gyr in age, and add this in quadrature to the formal uncertainties from our fit, which were $\pm$0.027 $M_{\odot}$ and $\pm$0.034 Gyr. In the case of $\rho_{s}$, which is not computed in \texttt{kiauhoku}, we propagate the final uncertainties from $R_{s}$ and $M_{s}$.

From this stellar isochrone fitting, we estimate that TOI-6054 has a mass of $1.107 \pm 0.040$ $M_{\odot}$ and an age of $6.0 \pm 1.1$ Gyr. As indicated by the best fit stellar evolution track (Figure \ref{fig:sed}, right panel), TOI-6054 is in the process of leaving the main sequence and is ascending the sub-giant branch. This unique position on the HR diagram is what allows us to localize the star's age well. It is rapidly moving along the HR-diagram as it expands, and will soon begin to cool off before becoming a red giant in likely the next $\sim$0.5 Gyr.

\subsection{Orbital and Planetary Parameters} \label{sec:orbpar}

We next fit the TESS and NEID observations of TOI-6054 to measure the orbital and planetary parameters of the two planets. We jointly fit the transit and radial velocity data. We used \texttt{batman} \citep{batman} to generate model transit lightcurves, and our own implementation of an analytic RV model which uses a numerical root finder to solve Kepler's problem. For the transit fits we used fixed quadratic limb darkening coefficients from \cite{limb_dark}, which we report in Table \ref{tab:plpar}.

\begin{figure}
    \centering
    \includegraphics[width=\columnwidth]{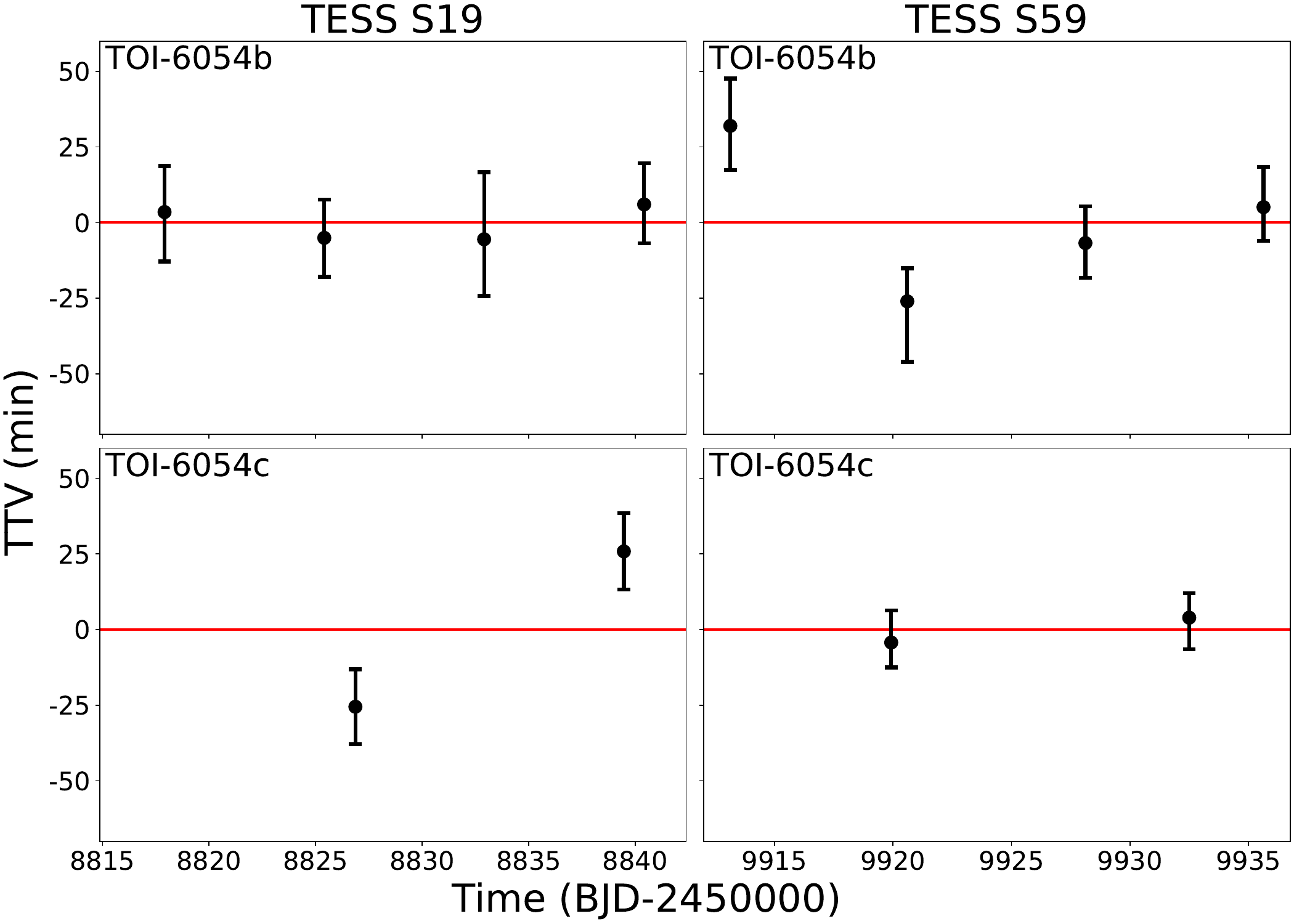}
    \caption{The transit timing variations between the best fit transit times and the linear ephemeris. We find that there is not significant evidence to include transit timing variations in our reported fit results. The top row shows the transit timing variations of TOI-6054b, while the bottom row shows those of TOI-6054c. The time axis is split into two columns for Sector 19 (left) and Sector 59 (right) for visual clarity.}
    \label{fig:ttvs}
\end{figure}

During our joint transit and RV fit, we fit for the following parameters for each planet: the log of the orbital period (log($P$)), the time of conjunction ($T_{C}$), the planet to star radius ratio ($R_{p}/R_{s}$), the log of the semi-major axis to stellar radius ratio (log($a/R_{s}$)), the cosine of the orbital inclination (cos($i$)), and the log of the RV semi-amplitude (log($K$)). Additionally, following \cite{exofast}, we parameterized the eccentricity ($e$) and argument of periastron ($\omega$) as $\sqrt{e}\text{cos}(\omega)$ and $\sqrt{e}\text{sin}(\omega)$.

We saw tentative evidence of transit timing variations (TTVs), so we performed a fit for each individual transit center time to test if TTVs are statistically preferred. We used a $\chi^{2}$ test to check whether the transit times deviate from the null hypothesis of a constant ephemeris. The resulting TTVs from this fit are plotted in Figure \ref{fig:ttvs}. The P-values of the $\chi^{2}$ tests are 0.310 and 0.032 for TOI-6054b and TOI-6054c respectively. TOI-6054b does not pass our significant P-value of 0.05, however TOI-6054c does marginally pass this test. However, with only four transits available, we believe that this could be an artifact of the small number of data points. We do not rule out the possibility of TTVs, but we choose not to fit them at this time.

\begin{figure*}[]
    \centering
    \includegraphics[width=\textwidth]{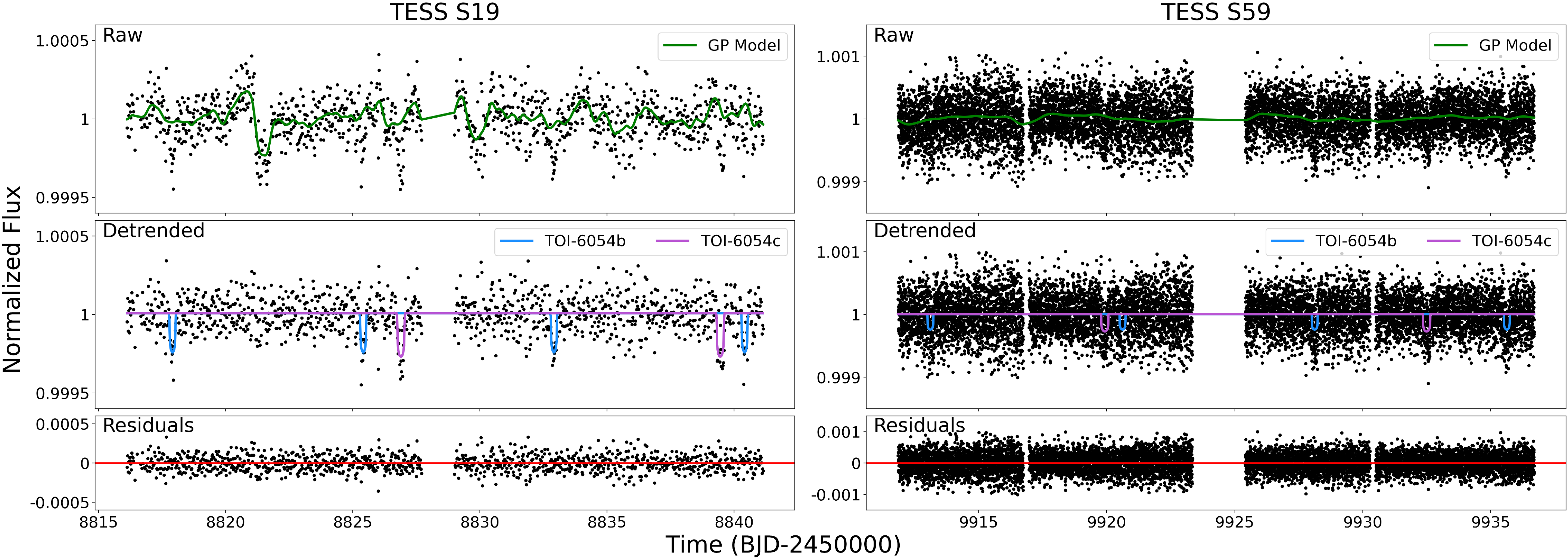}
    \caption{The fits to the TESS light curves, with Sector 19 on the left and Sector 59 on the right. {\it Top}: The raw PDCSAP SPOC data, and the best fit GP detrending model in green. {\it Middle}: The detrended data, with the best fit transit models for TOI-6054b and TOI-6054c plotted in blue and purple, respectively. {\it Bottom}: The residuals after subtracting the GP model and the transits.}
    \label{fig:det}
\end{figure*}

\begin{figure*}
    \centering
    \includegraphics[width=\textwidth]{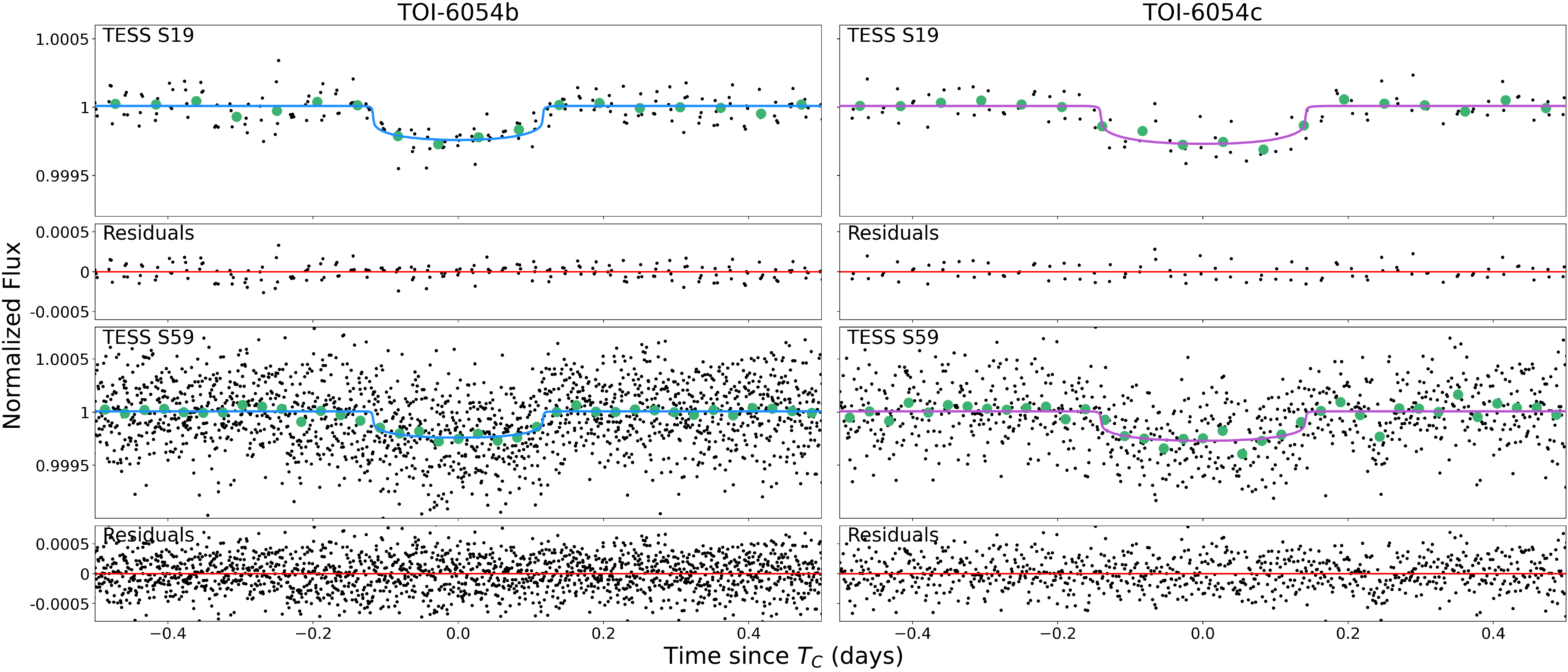}
    \caption{The phase-folded best fit transits models to the TESS data. The upper two rows are Sector 19 data and their fit residuals, and the lower two are the Sector 59 data and their fit residuals. The left column is the TOI-6054b transit model, and the right column is the TOI-6054c transit model.}
    \label{fig:transits}
\end{figure*}

Visual inspection of the TESS lightcurves (Figure \ref{fig:det}) showed evidence for stellar variability -- particularly in the longer cadence Sector 19 data. We fit this variability using a Gaussian Process regression. We used the \texttt{celerite2} \texttt{SHOTerm} model, which represents a stochastically driven, damped harmonic oscillator and we choose the commonly used value $1/\sqrt{2}$ for the quality factor (e.g., \citealt{celerite1}). For each sector, we fit for the log of the undamped period of the oscillator (log($\rho_{GP}$)) and the log of the standard deviation of the process (log($\sigma_{GP}$)), as well as a constant floating normalization value.

We tested different order polynomial background trends for the RV fit (0th-, 1st- and 2nd- orders).  We ran a set of test RV-only fits for each order, then compared the Bayesian Information Criterion (BIC) values of each of these fits. We found that the 0th-order background was preferred over the 1st-order and 2nd-order by about 3 points and 6 points, respectively. We therefore used a 0th-order RV background in all subsequent fits.

\begin{figure*}
    \centering
    \includegraphics[width=\textwidth]{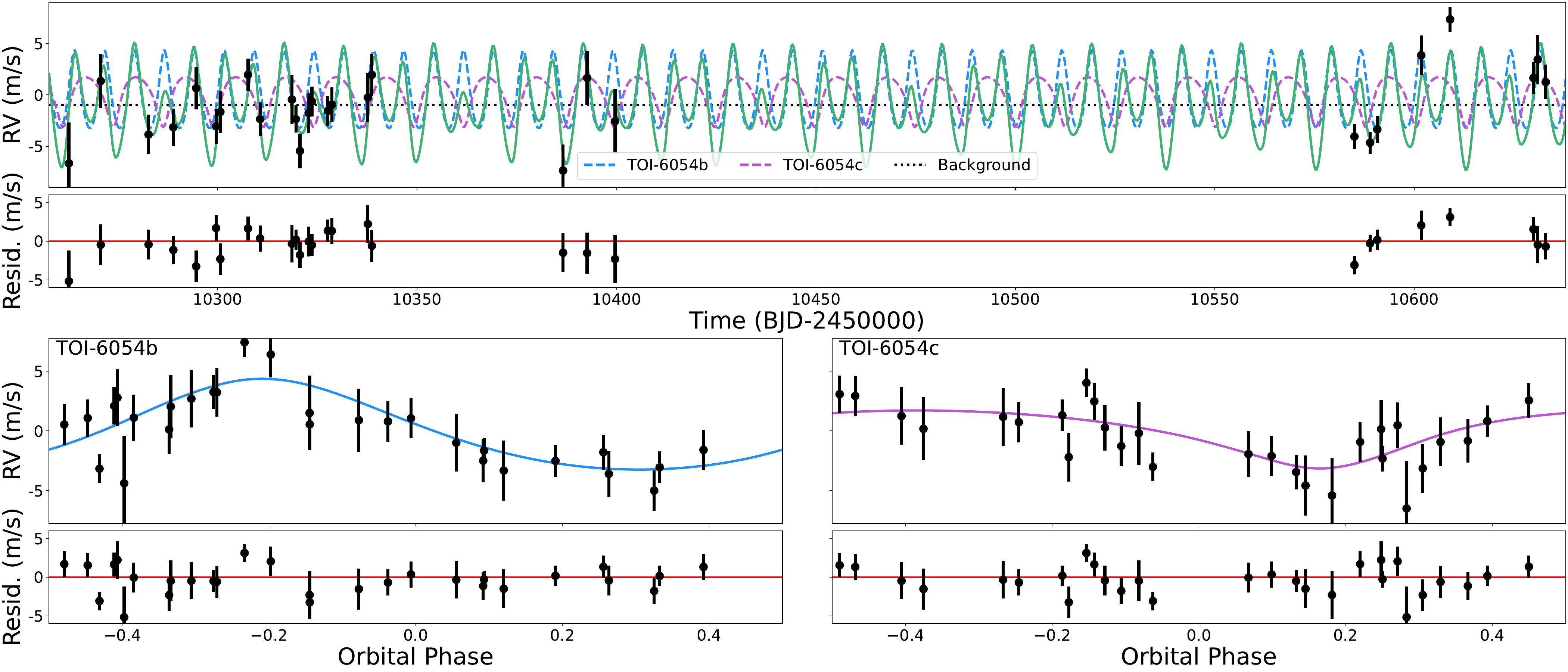}
    \caption{{\it Top}: The RV data are shown vs time in black, with the best fit combined RV model shown in green. The RV contribution from TOI-6054b is plotted as a blue dashed line, the contribution from TOI-6054c is plotted as a purple dashed line, and the contribution from the background trend is plotted as a black dotted line. Residuals are plotted below. {\it Lower Left}: The phase folded best fit RV model for TOI-6054b. The data plotted have had the background and TOI-6054c contributions subtracted. Residuals are plotted below. {\it Lower Right}: The phase folded best fit RV model for TOI-6054c. The data plotted have had the background and TOI-6054b contributions subtracted. Residuals are plotted below.}
    \label{fig:rv}
\end{figure*}

In total, our fit consisted of 23 parameters, and we scaled the errors in the transit and RV data to match the standard deviation of the residuals of our initial fits (these scaling factors are reported in Table \ref{tab:plpar}). We performed two version of our fit: one with uniform priors on all parameters and one with Gaussian priors on the transit stellar density from each planet ($\rho_{s,t}$), given by
\begin{equation}
    \frac{\rho_{s,t}}{g/cm^{3}} = 0.01891566 \left(\frac{a}{R_{s}}\right)^{3}\left(\frac{P}{1\text{ day}}\right)^{-2}
\end{equation}
where the coefficient absorbs the units and $P$ is the orbital period. The mean and width of the Gaussian priors for both planets was taken to be the value and uncertainty of the isochrone fitting results for TOI-6054's average density (see Table \ref{tab:stpar}). The uniform prior version of the fit was used to test that the stellar densities given by the transits are consistent with the measured value, and are further discussed in Section \ref{sec:fpa_rho}. All other results discussed elsewhere in this paper are referring to the fit version with the Gaussian priors on the transit stellar density.

We began our fits with an initial set of guess values taken from ExoFOP \citep{exofop}. We then used a Nelder-Mead minimizer on the likelihood function to improve our initial parameter estimates. We then used this initial best fit to perform a round of 5-sigma clipping on the transit light curves. During this process, we clipped 2 points from the Sector 19 data, and 25 from the Sector 59 data.

We then conducted an MCMC exploration of parameter space using the \texttt{emcee} Python package \citep{emcee}. We used 48 walkers with 10,000 burn-in steps and 20,000 sample steps, for a total of 960,000 samples after burn-in. We drew the initial walker positions from small Gaussian distributions centered on the parameter estimates from the Nelder-Mead minimizer. To confirm that the chains converged, we calculated the Gelman-Rubin statistic for each parameter \citep[][Chapter 13]{Gelman2013}. We found that the Gelman-Rubin statistic for all parameters was below the 1.1 threshold, therefore we concluded that the fit had converged. The standard deviation of the final fit residuals was 102 ppm in the Sector 19 light curve, 301 ppm in the Sector 59 light curve, and 1.8 m/s in the RV data.

\subsection{Results} \label{sec:res}

\begin{table*}[]
\centering
\caption{Joint transit and radial velocity fit results, reporting the median values and 68\% confidence intervals for each parameter.}
    \begin{tabular}{llll}
        \hline
        Parameter & Description & Value \\
        \hline
        Input Parameters \\
        \hline
        $u_{1}$ & 1$^{st}$ Order Limb Darkening Coeff.$^{1}$ & $0.3552$ \\
        $u_{2}$ & 2$^{nd}$ Order Limb Darkening Coeff.$^{1}$ & $0.2136$ \\
        $\sigma_{transit}$ & Transit Error Scaling Factor & Sector 19 & Sector 59 \\
        & & $1.3679$ &  $1.3706$ \\
        $\sigma_{RV}$ & RV Error Scaling Factor & $0.9950$ \\
        \hline
        Fitted Parameters & & TOI-6054b & TOI-6054c \\ 
        \hline
        log($P$) & Log Orbital Period (days) & $2.0150946_{-0.0000085}^{+0.0000076}$ & $2.5308066_{-0.000010}^{+0.0000094}$\\ 
        $T_{C}$ & Transit Time (BJD-2450000) & $8817.9087_{-0.0058}^{+0.0061}$ & $8826.8928_{-0.0091}^{+0.0097}$\\ 
        $R_{p}/R_{s}$ & Planet to Star Radius Ratio & $0.01464_{-0.00052}^{+0.00052}$ & $0.01552_{-0.00069}^{+0.00070}$\\ 
        log($a/R_{s}$) & Log Semi-Major Axis in Stellar Radii & $2.291_{-0.047}^{+0.043}$ & $2.636_{-0.048}^{+0.043}$\\ 
        cos($i$) & Cosine of Inclination & $0.026_{-0.017}^{+0.018}$ & $0.021_{-0.014}^{+0.014}$\\ 
        log($K$) & Log RV Semi-Amplitude (m/s) & $1.35_{-0.14}^{+0.13}$ & $0.93_{-0.25}^{+0.20}$\\ 
        $\sqrt{e}\text{cos}(\omega)$ & ... & $0.366_{-0.092}^{+0.064}$ & $-0.532_{-0.062}^{+0.096}$\\ 
        $\sqrt{e}\text{sin}(\omega)$ & ... & $-0.07_{-0.16}^{+0.17}$ & $-0.17_{-0.15}^{+0.13}$\\ 
        \hline
        Derived Parameters & & TOI-6054b & TOI-6054c\\ 
        \hline
        $P$ & Orbital Period (days) & $7.501437_{-0.000064}^{+0.000057}$ & $12.56364_{-0.00013}^{+0.00012}$\\ 
        $R_{p}$ & Radius ($R_{\oplus}$) & $2.65_{-0.15}^{+0.15}$ & $2.81_{-0.17}^{+0.18}$\\ 
        $a/R_{s}$ & Semi-Major Axis in Stellar Radii & $9.88_{-0.46}^{+0.44}$ & $13.96_{-0.65}^{+0.61}$\\ 
        $a$ & Semi-Major Axis (AU) & $0.0763_{-0.0048}^{+0.0048}$ & $0.1077_{-0.0067}^{+0.0067}$\\ 
        $i$ & Inclination (Degrees) & $88.51_{-1.0}^{+0.99}$ & $88.79_{-0.83}^{+0.82}$\\ 
        $b$ & Impact Parameter & $0.26_{-0.17}^{+0.19}$ & $0.29_{-0.20}^{+0.23}$\\ 
        $K$ & RV Semi-Amplitude (m/s) & $3.86_{-0.51}^{+0.51}$ & $2.53_{-0.55}^{+0.55}$\\ 
        $M_{p}$ & Mass ($M_{\oplus}$) & $12.4_{-1.6}^{+1.7}$ & $9.2_{-2.0}^{+2.0}$\\ 
        $e$ & Eccentricity & $0.163_{-0.059}^{+0.054}$ & $0.331_{-0.083}^{+0.072}$\\ 
        $\omega$ & Argument of Periastron (Degrees) & $-11_{-23}^{+27}$ & $-163_{-13}^{+16}$\\ 
        $\delta$ & Transit Depth (ppm) & $214_{-15}^{+15}$ & $241_{-21}^{+22}$\\ 
        $T_{14}$ & Total Transit Duration (hours) & $5.63_{-0.35}^{+0.33}$ & $6.62_{-0.58}^{+0.43}$\\ 
        $\rho_{p}$ & Density (g/cm$^{3}$) & $3.65_{-0.71}^{+0.86}$ & $2.27_{-0.60}^{+0.73}$\\ 
        $T_{eq}$ & Equilibrium Temperature (K) & $1360_{-31}^{+35}$ & $1144_{-26}^{+29}$\\ 
        $S_{inc}$ & Insolation Flux ($S_{\oplus}$) & $571_{-50}^{+60}$ & $286_{-25}^{+30}$\\ 
        $TSM$ & Transmission Spectroscopy Metric$^{2}$ & $37.6_{-6.1}^{+7.7}$ & $51_{-11}^{+17}$\\ 
        $\rho_{s,t}$ & Transit Stellar Density$^{3}$ (g/cm$^{3}$) & $0.324_{-0.043}^{+0.045}$ & $0.326_{-0.044}^{+0.044}$\\ 
        \hline
        Other Fit Parameters \\ 
        \hline
        $\gamma$ & RV Offset (m/s) & $-0.96_{-0.33}^{+0.34}$\\ 
        Transit Background & & Sector 19 & Sector 59 \\ 
        F$_{0}$ & Baseline Flux - $10^{6}$ (ppm) & $8_{-11}^{+10}$ & $7.6_{-9.6}^{+9.5}$\\ 
        log($\rho_{GP}$) & Log GP Period (days) & $0.07_{-0.17}^{+0.16}$ & $0.75_{-0.31}^{+0.32}$\\ 
        log($\sigma_{GP}$) & Log GP Std. & $-9.555_{-0.097}^{+0.10}$ & $-10.01_{-0.15}^{+0.15}$\\ 
        \hline
    \end{tabular}
    \label{tab:plpar}
    \begin{tablenotes}
        \item $^{1}$ \cite{limb_dark}. $^{2}$ Assuming zero albedo. $^{3}$ \cite{TSM}. $^{4}$ Gaussian priors were placed on the stellar density during fitting, using the measured value and uncertainty from stellar isochrone fitting (reported in Table \ref{tab:stpar}). The results reported in this table reflect recovery of that prior.
    \end{tablenotes}
\end{table*}

From our joint transit and radial velocity fit, we found that both planets are hot sub-Neptunes near the edge of the sub-Neptune desert \citep{desert,FP18}. TOI-6054b has an orbital period of $7.50$ days, a radius of $2.65_{-0.15}^{+0.15}$ $R_{\oplus}$, a mass of $12.4_{-1.6}^{+1.7}$ $M_{\oplus}$, and an equilibrium temperature of $1360_{-31}^{+35}$ K. TOI-6054c has an orbital period of $12.56$ days, a radius of $2.81_{-0.17}^{+0.18}$ $R_{\oplus}$, a mass of $9.2_{-2.0}^{+2.0}$ $M_{\oplus}$, and an equilibrium temperature of $1144_{-26}^{+29}$ K.

These results are consistent with the ``peas in a pod" expectations from \cite{Weiss2018}. TOI-6054b and c are very similar in radii, with the cooler outer planet being slightly larger. Additionally, the two planets are separated by $\sim$13 mutual Hill radii, relatively tightly packed compared to most Kepler two-planet systems \cite{Weiss2018,Lissauer2011}.

We show the full results of our transit and radial velocity fitting in Table \ref{tab:plpar}, including derived parameters. Figures \ref{fig:det}, \ref{fig:transits}, and \ref{fig:rv} show the results of the GP light curve detrending, the phase-folded best fit transits, and the best fit RV model, respectively. Figure \ref{fig:corner} shows the full posterior distribution of our joint fit.

We note that there is modulation in the RV fit residuals that could be due to a non-transiting planetary component. To check for this, we used a $\chi^{2}$ test to determine whether the RV residuals deviate from the null hypothesis of zero. The P-value of this test was 0.421, which does not pass our significant P-value of 0.05. Therefore, we did not find sufficient evidence to fit for a third planetary component in our RV model. We do not rule out additional planetary components altogether, as the addition of more RV data may reveal non-transiting planets in the future.

\subsection{False Positive Analysis} \label{sec:fpa}

We conducted a set of false positive analyses to further confirm the planetary natures of TOI-6054b and TOI-6054c. We note that the phasing of the RV data to both of the photometric orbital periods and the lack of nearby sources in follow-up imaging very strongly indicate that there are two bona fide planets orbiting TOI-6054. Nevertheless, we wished to further investigate possible false positive scenarios.

This investigation was partially prompted by the fact that the orbital period of TOI-6054c is close to the likely rotation period of the star TOI-6054 itself. Combining the measured spectroscopic $v$sin $i_{s}$ with our SED-derived estimate of $R_{s}$, we estimate a stellar rotation period of $12.6 \pm 1.1$ days -- assuming the stellar rotation axis is in the plane of the sky -- although we do not find this rotation signal in the TESS data. This is suggestively close to the $12.56364 \pm 0.00013$ day orbital period of TOI-6054c. However, as we describe below, we believe this is coincidental.

First and foremost, if TOI-6054c were a starspot or other stellar surface feature, we would not expect to see a transit-like signal at the same depth, duration, phase, and period across the two TESS sectors of data, which are separated by three years. Second, we measured consistent stellar densities from the transits of both planets when we placed uniform priors on both. Finally, we do not detect any correlation between the RV signals and the spectral features of the star.

\begin{figure}
    \centering
    \includegraphics[width=\columnwidth]{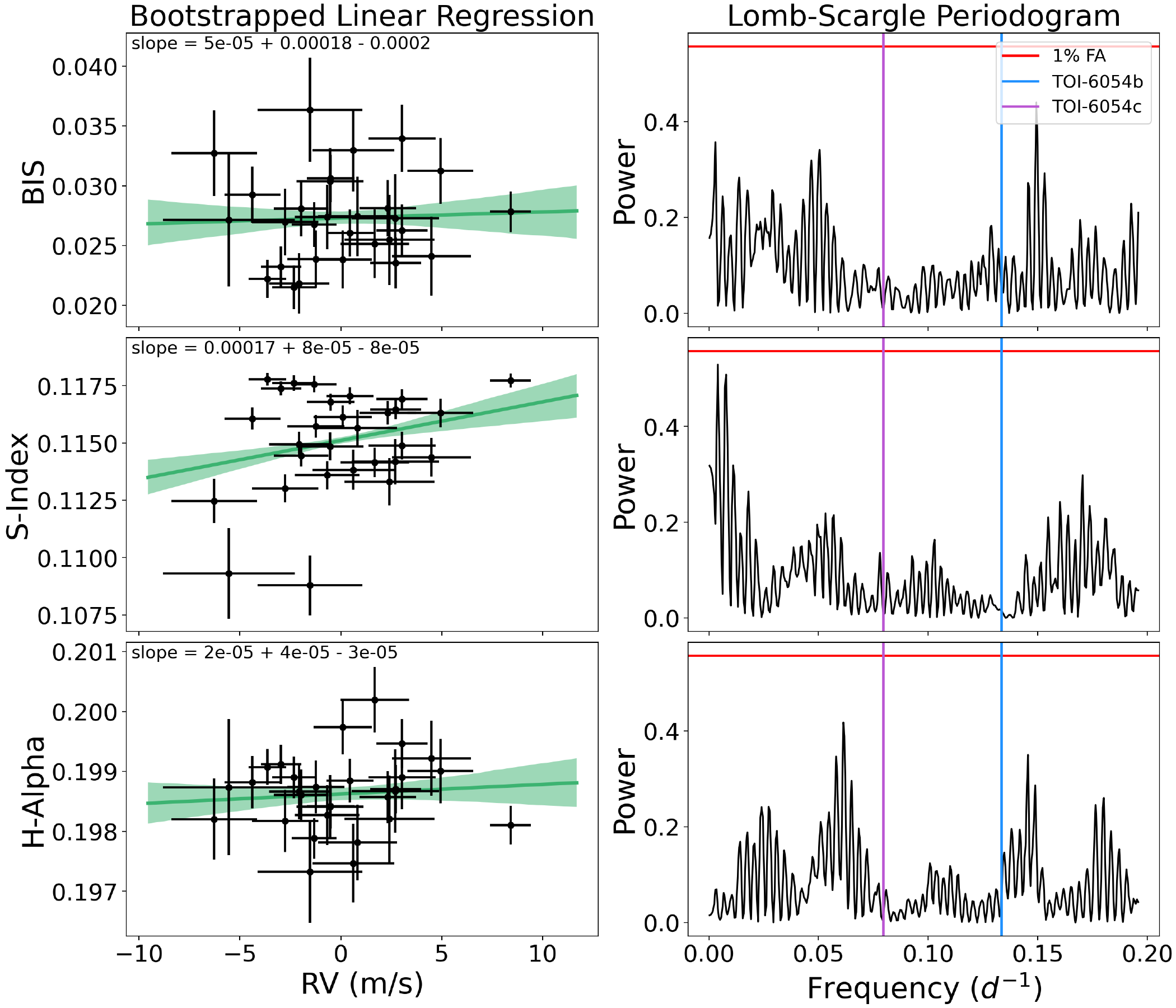}
    \caption{{\it Left Column}: BIS or stellar activity indices vs RV. The data are shown in black and the best fit bootstrapped linear regression model is shown in green with a 1$\sigma$ envelope. The best fit slope and its uncertainties are printed at the top of each plot. {\it Right Column}: Lomb-Scargle periodograms, with the orbital frequencies of TOI-6054b and TOI-6054c shown in blue and purple respectively. The 1\% false alarm value is in red. The top row is the BIS, the middle row is the S-Index, and the bottom row is H$\alpha$-index.}
    \label{fig:bis_act}
\end{figure}

\subsubsection{Stellar Density} \label{sec:fpa_rho}

One check on the planetary-nature of TOI-6054b and TOI-6054c is the host star density from the transit fits. The stellar density from our stellar isochrone analysis of TOI-6054 is $0.340 \pm 0.045$ g/cm$^{3}$ (Section \ref{sec:iso}). The host star densities from our fits to the transit lightcurves, with uniform priors placed on these values (see Section \ref{sec:orbpar}), are $0.13_{-0.10}^{+0.16}$ and $0.21_{-0.14}^{+0.22}$ g/cm$^{3}$ for TOI-6054b and TOI-6054c, respectively. Both of the transit stellar densities are consistent with the stellar isochrone analysis at near 1$\sigma$, and, importantly, with each other.

\subsubsection{Line Bisector Analysis} \label{sec:fpa_bis}

To check if the signals from TOI-6054b and TOI-6054c were caused by a blended or nearby eclipsing binary not detected in follow-up imaging, we performed an analysis of the spectroscopic line bisectors in our NEID observations. We used the bisector inverse slope (BIS) from the NEID Level 2 data products, which is the velocity difference between the top and bottom portions of the CCF bisector, as described in \cite{NEID_DRP}. We checked for any correlation of these values with the RV measurements using three methods: a bootstrapped linear regression, a Lomb-Scargle periodogram, and the Pearson correlation coefficient (PCC). If one of the planetary signals is caused by an eclipsing binary, we expect to see a strong correlation between the RV measurements and the BIS values due to line blending \citep{blend}.

For our bootstrap estimation of correlations in the line bisectors, we drew 1000 random samples for each BIS vs RV data point. We drew the samples from a bivariate normal distribution centered on the measured values, with widths corresponding to the uncertainties on the BIS and RV measurements. We then performed a linear regression -- without accounting for measurement uncertainties -- on each of the 1000 draws. We took the best fit slope to be the median of the draws, and we used the 68\% confidence interval of the draws for the uncertainties on this slope. The slope of the best fit line was $0.00005 \pm 0.00019$, consistent with 0 within 1$\sigma$, which indicates that there is no apparent correlation between the line bisectors and RV measurements (Figure \ref{fig:bis_act}, top left panel).

We also performed a Lomb-Scargle analysis on the line bisectors to check for any periodicities. We do not see any significant peaks near the planetary orbital periods (Figure \ref{fig:bis}, top right panel). This is another indicator that there is no correlation between the line bisectors and the RVs.

Finally, we calculated the PCC for the BIS vs RV data. The resulting PCC value was $0.04$, with a two-sided p-value of $0.838$. In other words, random noise would produce a PCC value of this magnitude, or greater, 83.8\% of the time. The standard p-value for a significant correlation is $0.05$. Once again, this indicates no significant correlations exist between the BIS values and the RV data. Taking into account these analyses, we can rule out eclipsing binaries as causing the RV signal.

\subsubsection{Stellar Activity Indicators} \label{sec:fpa_act}

We also investigated possible correlations between stellar activity indicators and the RV data. NEID Level 2 data products contain measurements of multiple stellar activity indicators, as described in \cite{NEID_DRP}. We chose to use the S-index and H$\alpha$-index as they are widely used stellar activity indicators \citep{act_ind}. We use the telluric line corrected stellar activity indicators provided in the NEID Level 2 data products, as they show greater long term stability \citep{NEID_DRP}. Once again, we checked for correlations using a bootstrapped linear regression, a Lomb-Scargle periodogram, and a PCC calculation. We follow the same methods for all three as in Section \ref{sec:fpa_bis}.

\begin{figure*}
    \centering
    \includegraphics[width=\textwidth]{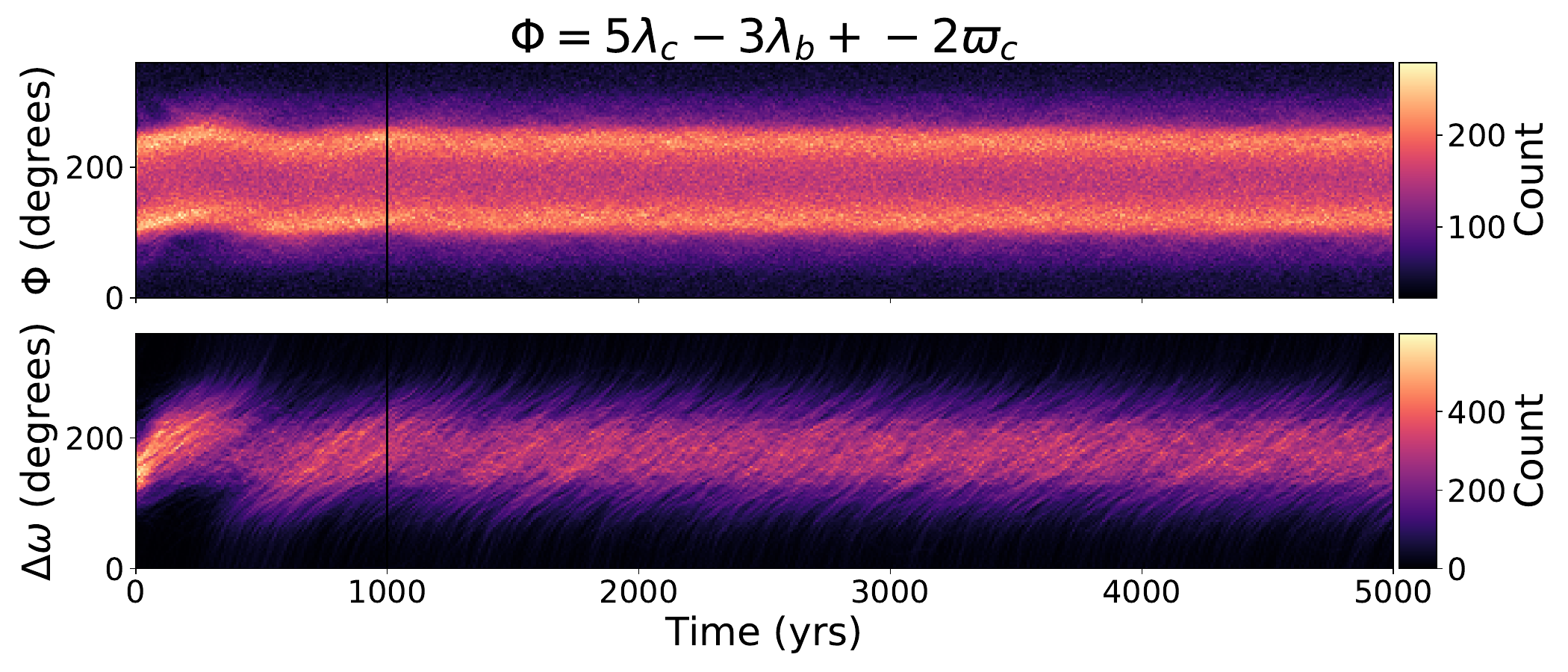}
    \caption{{\it Top}: Density plot of the resonant argument given at the top of the plot vs simulation time, with the color scale showing the number of simulations in each bin. The initial $\sim$800 year settling phase is marked with a black vertical line. {\it Bottom}: Density plot of the difference between the arguments of periastron of the two planets vs simulation time, with the color scale showing the number of simulations in each bin. The diffusion period is marked in this plot as well.}
    \label{fig:res}
\end{figure*}

After performing the bootstrapped linear regression on the S-Index vs RV data, we found a slope of $0.00017 \pm 0.00008$, which is somewhat ($\sim$2$\sigma$) inconsistent with a value of zero (see Figure \ref{fig:bis_act}, middle left panel). However we note that this is heavily biased by a single low S-Index value with large uncertainty seen in the lower left corner of the plot, associated with our very first RV observation. We also found that the values and uncertainties of the S-Index are highly correlated, having a PCC value of $-0.887$ with a p-value of $1.6\times10^{-10}$. We do not see significant peaks at either planet's orbital frequency in the Lomb-Scargle periodogram of the S-Index data (Figure \ref{fig:bis_act}, middle right panel). The PCC value of the S-Index vs RV data is $0.306$ with a p-value of $0.106$, which is above the standard significant p-value of $0.05$. We therefore do not find significant evidence for a correlation between the S-Index and RV measurements in our observations.

We performed the same analysis on the H$\alpha$-index. We found a slope of $0.00002 \pm 0.00004$ using the bootstrap regression, consistent with 0 within 1$\sigma$ (Figure \ref{fig:bis_act}, lower left panel). We do not see significant peaks in the Lomb-Scargle periodogram of the H$\alpha$-index at the orbital frequencies of either planet (Figure \ref{fig:bis_act}, lower right panel). The PCC value of the H$\alpha$-index vs RV data is $0.107$ with a p-value of $0.58$, well above $0.05$. We therefore find that there is no significant correlation between the H$\alpha$ and RV measurements. The lack of correlation between the RV data and the stellar activity indicators provides further evidence that TOI-6054b and TOI-6054c are planetary in nature.

\section{Discussion} \label{sec:dis}

\subsection{Mean Motion Resonance} \label{sec:dyn_mmr}

Due to the period ratio of the two planets being very near 5:3, we ran a check for mean motion resonance in the system. To do this, we used the \texttt{REBOUND} N-body python package \citep{rebound}, and integrated the simulations using WHFast, a symplectic Wisdom-Holman integrator \citep{reboundwhfast,wh}.

We simulated the planetary orbital dynamics by randomly drawing 2,000 MCMC samples from the joint transit and RV fit, using the orbital parameters and planetary masses in each to initialize the simulations. The longitude of the ascending node was initially set to zero for both planets. In each simulation, we integrated the system forward in time for 5000 years (greater than 200,000 orbits of the inner planet) with a time-step of 0.075 days, which is equal to 1\% of the period of TOI-6054b, chosen to achieve fractional energy changes to better than one part in 10$^{6}$. We threw out any simulation in which there were any collisions, the semi-major axis of either planet exceeded 10 AU, or the fractional energy change exceeded $10^{-6}$ (778 of the 2000 simulations). In the rest, we tracked the argument of periastron, the longitude of periastron ($\varpi=\omega + \Omega$), and the mean longitude ($\lambda=\varpi + M$, where $M$ is the mean anomaly) of both planets every year. We used these orbital elements to calculate resonant arguments for the two planets as
\begin{equation}
    \Phi = 5 \lambda_{c} - 3 \lambda_{b} + r \varpi_{c} + s \varpi_{b},
\end{equation}
where the subscripts $b$ and $c$ denote TOI-6054b and TOI-6054c, respectively, and $r$ and $s$ are small integers chosen such that the sum of all four coefficients is zero \citep[][Chapter 8]{Murray_Dermott}. We also computed the difference between the arguments of periastron $\Delta\omega = \omega_{b} - \omega_{c}$ as a function of time. We found that the resonant argument most commonly seen to librate in this system was $\Phi = 5 \lambda_{c} - 3 \lambda_{b} - 2 \varpi_{c}$ (this process is described later in this section). We show the density plots of this resonant argument and $\Delta\omega$ in Figure \ref{fig:res}.

As illustrated in Figure \ref{fig:res}, both $\Phi$ and $\Delta\omega$ librate. The resonant argument displays libration centered on roughly 180$^{\circ}$ with peaks around 105$^{\circ}$ and 255$^{\circ}$ in many of the simulations. We see $\Delta\omega$ librating in a slightly larger range. Additionally, both show an initial dynamical burn-in of approximately 1000 years, which is marked with a black vertical line on each. This occurs because the observations constrain these values to be near a unique phase in their oscillations, and the phases spread out over time.

We quantified the probability of the system being in the 5:3 mean motion resonance following the procedure of \cite{MMR}. For each simulation, we began by trimming the first 1000 years to avoid the dynamical burn-in, and we then computed 13 resonant arguments, iterating $r$ from -6 to 6. For each resonant argument, we used 36 angular bins (10$^{\circ}$ per bin) and 60 time bins (approximately 67 years per bin) to divide up the simulated values into a grid. We flagged all boxes with no points, and found the largest contiguous group of flagged boxes in each time bin. The median of the counts of all of the time bins was used as the resonance parameter for that simulation.

From visual inspection of individual simulations, we determined that those with a resonance parameter value of six or more were librating in $\Phi$, and are therefore in resonance. Using this as our criteria, we found that $\Phi = 5 \lambda_{c} - 3 \lambda_{b} - 2 \varpi_{c}$ displayed libration the most often, occurring in 92.64\% of the 1,222 remaining simulations. In total, 93.29\% of the simulations displayed libration in at least one resonant argument. The high fraction of librating cases in this sample indicates that the TOI-6054 system is likely very close to or within the 5:3 mean motion resonance.

\begin{figure*}
    \centering
    \includegraphics[width=\textwidth]{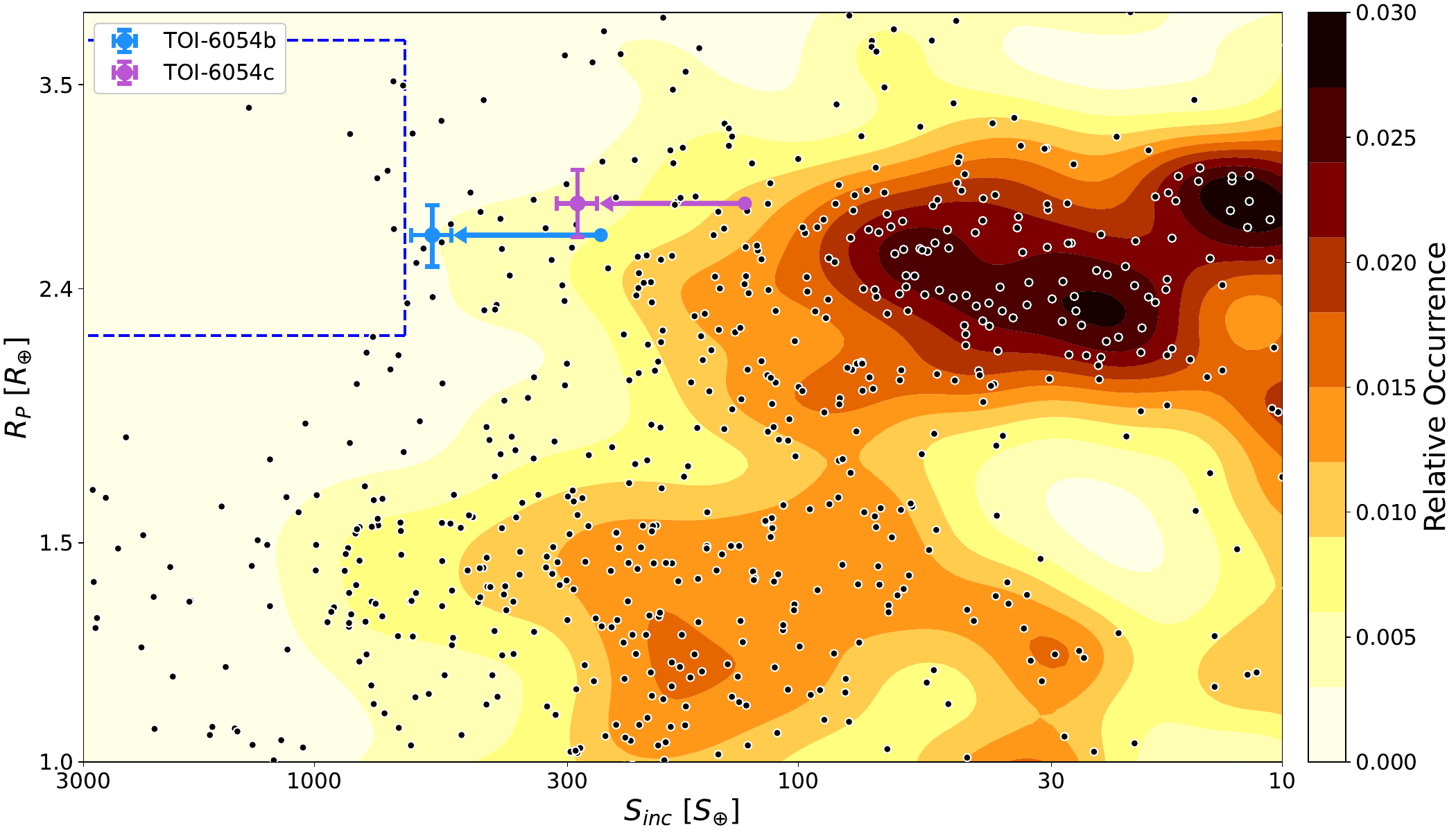}
    \caption{$R_{p}$ vs $S_{inc}$ for the California-Kepler Survey planets from \protect\cite{FP18} are plotted in black. The background color map is a weighted KDE of these planets showing relative occurrence rates as defined in that work. The blue dashed lines mark off the sub-Neptune desert defined in \protect\cite{desert}. TOI-6054b and TOI-6054c are over-plotted with error bars in blue and purple, respectively. The blue and purple dots with arrows depict the estimated ZAMS $S_{inc}$ for TOI-6054b and TOI-6054c, respectively, and their increase as the host star has evolved.}
    \label{fig:flux}
\end{figure*}

The likely mean motion resonance of the TOI-6054 system is consistent with the planets migrating during formation. Planet formation simulations show that inwards migration of protoplanets often results in groups of planets in stable mean motion resonances \citep{Cresswell2008,Batygin2015}. In the case of TOI-6054b and c, this likely would have been type I migration due to asymmetric torques exerted by the disk, as smaller protoplanets tend to execute this type of migration \citep{ward97}. This is also supported by the fact that the best fit $\Delta\omega$ between the two planets is $151_{-30}^{+31}$ degrees, consistent with 180 degrees at 1$\sigma$. \cite{batygin2013a} found that convergent migration preferentially results in anti-alignment of the arguments of periastron, as is seen in the TOI-6054 system.

Due to the TOI-6054 system's likely mean motion resonance, it is possible that the transit times of the planets exhibit a long term super-cycle. Unfortunately, with only eight and four transits across two TESS Sectors for TOI-6054b and c, respectively, we cannot constrain this cycle, or whether or not it even exits. However, with more data in the future, we may be able to see evidence of this cycle as well as refine the periods and masses of both planets. The TOI-6054 system is scheduled to be observed again in TESS Sector 86 from UT 2024 Nov 21 to UT 2024 Dec 18, and we plan to perform follow up in the future as well. With more than two epochs of transit observations, we may start to see a deviation from a linear ephemeris.

\subsection{Incident Flux} \label{sec:flux}

The zero-albedo black-body equilibrium temperatures of TOI-6054b and TOI-6054c, respectively, are $T_{eq}=1360_{-31}^{+35}$ K and $T_{eq}=1144_{-26}^{+29}$ K. The stellar flux incident upon TOI-6054b is $571_{-50}^{+60}$ $S_{\oplus}$, and on TOI-6054c it is $286_{-25}^{+30}$ $S_{\oplus}$. The incident flux upon TOI-6054b is relatively high for a sub-Neptune, as most planets of this size are found at much lower incident stellar flux values \citep{desert, FP18, radius_valley}.

As noted in Section \ref{sec:iso}, TOI-6054 is leaving the main sequence and evolving into a more luminous sub-giant. This implies that the incident flux on TOI-6054b and TOI-6054c has been increasing over time. Based on the isochrone analysis in Section \ref{sec:iso}, we estimate that TOI-6054 had a zero age main sequence (ZAMS) radius of 1.06 $R_{\odot}$ -- compared to its current radius of 1.662 $R_{\odot}$. Since the ZAMS, TOI-6054's luminosity has increased from 1.5 $L_{\odot}$ to its current value of 3.36 $L_{\odot}$. This has increased the amount of radiation received by TOI-6054b compared to when the planet formed, which allows it to temporarily exist above the radius valley near the sub-Neptune desert \citep{desert}. If TOI-6054b has retained its primordial H/He dominated atmosphere, we estimate that the planet could be currently losing it (see Sections \ref{sec:he} and \ref{sec:lyman}). Future atmospheric observations of TOI-6054b may provide a window into the processes thought to sculpt the radius valley \citep{radius_valley}.

\begin{figure*}
    \centering
    \includegraphics[width=\textwidth]{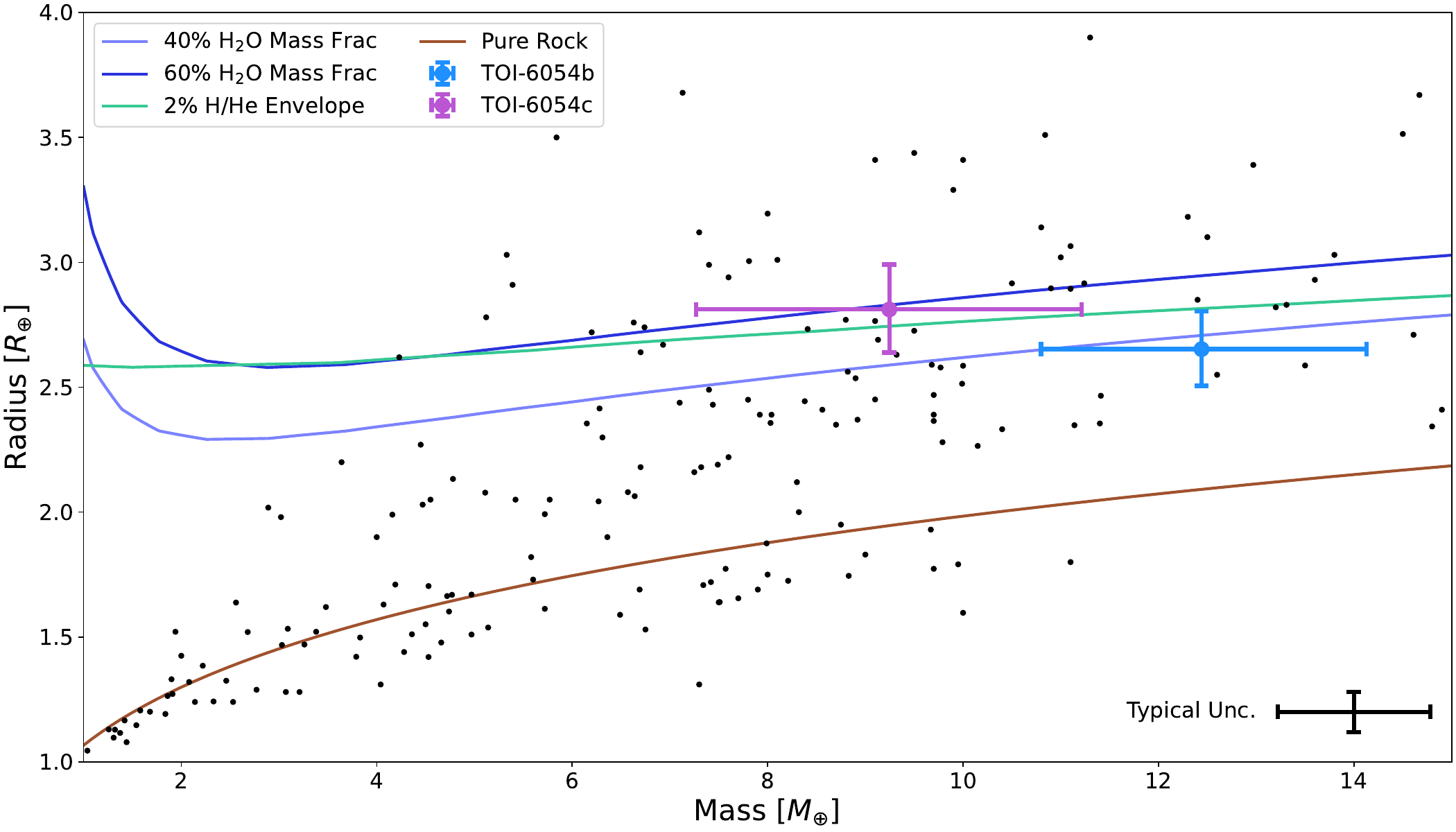}
    \caption{Mass-radius diagram of confirmed exoplanets from the NASA Exoplanet Archive smaller than 4 $R_{\oplus}$ and less massive than 15 $M_{\oplus}$ (black dots). We filter these for planets with $
    \leq20$\% mass uncertainties and $\leq10$\% radius uncertainties. TOI-6054b and TOI-6054c are over-plotted in blue and purple respectively. Mass-radius curves for water-world planets at 40\% and 60\% H$_{2}$O mass fractions, rocky worlds with a 2\% H/He envelope mass fraction and an Earth-like core composition, and pure silicate rocky worlds are plotted in blue, green, and brown, respectively. See section \ref{sec:comp} for the details of these curves.}
    \label{fig:mr}
\end{figure*}

In Figure \ref{fig:flux}, we have recreated the right panel of Figure 6 in \cite{FP18}, which shows occurrence rates of Kepler planets in $R_{p}-S_{inc}$ space. We also outline the sub-Neptune desert as defined in \citep{desert}. Including TOI-6054b and TOI-6054c, we see that the former is relatively unusual in $R_{p}-S_{inc}$ space for sub-Neptunes above the radius valley. In fact, the median fit values for TOI-6054b place it just outside of the sub-Neptune desert. However, we estimate that TOI-6054b and TOI-6054c had incident fluxes of roughly 256 $S_{\oplus}$ and 129 $S_{\oplus}$, respectively, when TOI-6054 was a ZAMS star (plotted as dots with arrows leading to their current values), so the planets have been slowly pushed towards the sub-Neptune desert over time as the stellar flux has increased. Although still relatively high, these values are much more in line with expectations for sub-Neptunes.

\subsection{Composition} \label{sec:comp}

We measure densities of $3.65 \pm 0.79$ g/cm$^{3}$ for TOI-6054b, and $2.27 \pm 0.67$ g/cm$^{3}$ for TOI-6054c. Placing these planets on a mass-radius diagram (Figure \ref{fig:mr}), we can see that their radii are much too large to be pure rocky planets. Rather, they are quite similar to many other sub-Neptunes in mass.

As both planets in this system are relatively highly irradiated, we followed a similar approach to \cite{TOI-544}, which reported the discovery of another hot sub-Neptune, in order to estimate bulk compositions. We used the \texttt{Structure Model INTerpolator (smint)} python package \citep{smint} to interpolate onto planet composition model grids, and we obtained posterior distributions on the compositions via MCMC. We performed this for two potential compositions: rocky worlds with H/He envelopes and irradiated water-worlds with no atmosphere.

When fitting for H/He envelope compositions, we had the option between the model grids of \cite{LF14} and \cite{Zeng19}. However, the \cite{Zeng19} H/He mass-radius models assume constant specific entropy, which is not accurate for planets that have undergone thermodynamic processes, such as cooling and mass-loss, which reduce their specific entropy \citep{Rogers23}. We instead used the \cite{LF14} models, which account for the thermal evolution of planets \citep{Rogers23}. Figure 7 of \cite{TOI-544} provides an excellent comparison of these two model grids. Fitting for the H/He envelope planet composition using the \cite{LF14} model grids at 50 times Solar metallicity with an Earth-like core composition, we obtained an H/He envelope mass fraction of 1.39\% for TOI-6054b and an envelope mass fraction of 2.52\% for TOI-6054c.

When fitting water-world composition models, we had the option between the \cite{Zeng19} and \cite{Aguichine21} model grids. We used the latter, as they were specifically computed for highly-irradiated planets and use a more correct equation of state for water. We estimated an H$_{2}$O mass fraction of 37\% for TOI-6054b and 60\% for TOI-6054c using these models. Considering the very high water mass fractions fit with this model, we conclude that these planets likely have either a primary H/He atmosphere or a heavier secondary atmosphere.

There exists a strong degeneracy between H/He envelope compositions and water-world compositions for sub-Neptune planets which cannot be broken using only mass and radius measurements \citep{Rogers23}. Realistically, the TOI-6054 planets are most likely not H/He enveloped rocky worlds or H/He-free water-worlds; they are likely intermediate between these two extremes, as many sub-Neptunes are expected to have formed beyond the water ice-line in their systems \citep[e.g.,][]{Venturini20, hycean}. As we describe in Section \ref{sec:atmo}, atmospheric transmission spectroscopy or atmospheric escape observations would be able to break this degeneracy.

In Figure \ref{fig:mr}, we have plotted TOI-6054b and TOI-6054c on a mass radius diagram of small ($R_{p} \leq 4\:R_{\oplus}$, $M_{p} \leq 15\:M_{\oplus}$) confirmed exoplanets from the NASA Exoplanet Archive. Over-plotted are a set of representative mass-radius curves for varying H/He envelope (at $S_{inc}=428\:S_{\oplus}$) and H$_{2}$O mass fractions (at $T_{eq}=1252$ K). Additionally, we plot a mass radius curve for pure silicate rocky worlds from \cite{Zeng19}, to further highlight that these planets cannot be pure rock. The strong degeneracy between the H/He envelope curves and the water-world curves is very clear here, especially taking into account the error bars on the TOI-6054 points.

\section{Potential for Atmospheric Characterization} \label{sec:atmo}

The TOI-6054 system has the potential to be an excellent candidate for future atmospheric follow-up observations. With two similarly sized sub-Neptunes around a bright star, atmospheric composition constraints of both planets could provide significant information about their formation and evolution, including breaking the compositional degeneracy discussed in Section \ref{sec:comp}. Additionally, as described in Section \ref{sec:flux}, TOI-6054b could be losing its primordial H/He atmosphere if it is still present. If this is indeed the case, we expect to be able to measure a strong helium triplet signal around 1083 nm \citep{he_line}. This feature has been shown to be highly dependent on atmospheric metallicity \citep{he_escape}, and thus can be used to constrain the metallicity in atmospheric retrievals, improving our ability to detect other features. We also expect to be able to observe Lyman $\alpha$ transits of TOI-6054b \citep{lymanalpha}.

\subsection{JWST Transmission Spectroscopy}\label{sec:jwst}

The brightness of the star TOI-6054 and the two similar mass, sub-Neptune, planets make the system an attractive target for follow-on atmospheric characterization observations with JWST. Primarily because of the brightness of TOI-6054, the transmission spectroscopy metric \citep[TSM,][]{TSM} is $37.6{-6.1}^{+7.7}$ for TOI-6054b and $51_{-11}^{+17}$ for TOI-6054c, indicating both planets could show detectable atmospheric features in transmission spectroscopy observations. Additionally, the relatively high 1359\,K equilibrium temperature for TOI-6054b places it in a region of the $R_p$ vs. $T_{eq}$ parameter space that has very few other known sub-Neptunes with masses measured to better than $5\,\sigma$. However, as noted in \cite{TSM}, the standard TSM calculation is not meant for use with planets around bright ($J<9$) host stars due to JWST's lower duty cycle for bright star observations. 

The only JWST instrument that can observe the TOI-6054 system without saturating is NIRCam. We simulated timeseries observations of both planets' transmission spectra using NIRCam grism spectroscopy in the F322W2 and F444W filters and observations using the newly commissioned NIRCam DHS spectroscopic mode using \texttt{GenTSO} \citep{Cubillos2024_GenTSO}. The typical observing duty cycle for all of the simulations was approximately 80\%. The simulated F322W2 and F444W data had predicted transit depth uncertainties of roughly $\pm20$\,ppm in a single transit at $R=100$, while the simulated DHS data had uncertainties of $\pm40$\,ppm at $R=100$ in a single transit.

We then used \texttt{ExoTransmit} \citep{Kempton2017_ExoTransmit} to generate transmission spectra for both planets assuming a cloud-free, moderately super-solar metallicity atmosphere of M/H=100 in chemical equilibrium. The predicted transmission spectra for both planets' were effectively the same. The major molecular features at the equilibrium temperatures of the TOI-6054 planets that NIRCam DHS, F322W2, and F444W cover are \water, \methane, and \carbondiox. The largest feature predicted in a NIRCam transmission spectrum is the \carbondiox\ absorption near 4.5\um, which our modeling predicted to have an amplitude of about 35\,ppm above the continuum. The broad \water\ feature just shortwards of 3\um\ is expected to have an amplitude of about 25\,ppm, and the \water\ and \methane\ features visible in the shorter wavelength DHS data are all roughly 10\,ppm in amplitude.

To observe either planets' atmosphere and detect features at high significance would therefore require several (two to four) transits on each planet. Since the transit durations for both planets are relatively long, at about 6 hours each, such a JWST observing program would require 30 to 60 hours of charged time on JWST for each planet. We do note that based on the analysis done by \cite{Brande2024_CloudTrends}, we would expect that both TOI-6054b and TOI-6054c will show relatively cloud-free atmospheres, since both planets have equilibrium temperatures significantly hotter than the ``maximum cloudiness'' range of 500\,K - 800\,K estimated in that work. 

\subsection{1083 nm Helium Triplet}\label{sec:he}

Although transmission spectroscopy of the TOI-6054 system is time-intensive, we expect the 1083 nm helium triplet to be observable from the ground. To estimate this, we rescaled the observed signal of TOI-836c from \cite{he_escape}, a sub-Neptune of similar size to TOI-6054b \citep{836detection}. We make use of their updated empirical correlation between the order-of-magnitude mass loss rate estimated from the equivalent width of helium absorption ($\dot{m}_{obs}$) and the energy-limited mass loss rate ($\dot{m}_{theory}$), first reported in \cite{he_relation}. $\dot{m}_{theory}$ is defined in that work to be
\begin{equation}\label{eq:mdotth}
    \dot{m}_{theory} = \frac{\pi (1.25 R_{p})^{3} F_{XUV}}{G M_{p}}
\end{equation}
where $F_{XUV}$ is the incoming flux in the 5-504 \AA\:range. Unfortunately, we do not have access to data in this full range, so we instead use SRG/eROSITA \citep{eROSITA} data in the 5-62 \AA\:range and rescale $\dot{m}_{theory}$ for TOI-836c.

TOI-836 is part of the SRG/eROSITA catalog, and, after rescaling the flux for the star's distance and the orbital distance of TOI-836c, we found $F_{XUV,836c}=402$ erg/s cm$^{2}$. TOI-6054 is not in the catalog however, so we cross matched SRG/eROSITA with the TESS Input Catalog version 8.2 \citep{tic8.2} and searched for a star similar to TOI-6054. We chose to use the star TIC 70227585 as our proxy for TOI-6054, as it has $T_{eff}=6045\pm 131$ K, log $g_{s}=4.04\pm0.08$, $R_{s}=1.671\pm0.08$ $R_{\odot}$, and $M_{s}=1.12\pm0.15$ $M_{\odot}$. Using the measured X-ray flux and rescaling for the distance of the star and the planetary semi-major axis, we found $F_{XUV,6054b}\simeq86000$ erg/s cm$^{2}$.

We then plugged these values into equation \ref{eq:mdotth}, and used the empirical relation of \cite{he_escape}
\begin{equation}
    \dot{m}_{obs}=0.29 \dot{m}_{theory}^{0.481}
\end{equation}
to find $\dot{m}_{obs}$ for each planet. We found that $\dot{m}_{obs,6054b}=0.47\:M_{\oplus}$/Gyr, and assuming an atmospheric mass fraction of 1.39\% (from Section \ref{sec:comp}) and a constant mass loss rate, we estimate that TOI-6054b would lose its entire atmosphere in about 350 Myr.

Noting that the equivalent width ($W_{avg}$) of the helium absorption signal is proportional to the transit depth ($\delta$), and that $\dot{m}_{obs}\propto W_{avg}$ \citep[][Eq 1]{he_relation}, we rescaled the measured $W_{avg}$ of TOI-836c for the TOI-6054 system. We rescaled the error on $W_{avg,836c}$ assuming it is due to pure photon noise in the J band. From this analysis, we estimate a helium triplet signal to noise of 30 for TOI-6054b. Compared to the measured signal to noise of TOI-836c of 11.2 \citep{he_escape}, we conclude that ground based observations of the helium triplet for TOI-6054b are very feasible.

\subsection{Lyman \texorpdfstring{$\alpha$}{Alpha}}\label{sec:lyman}

Assuming that TOI-6054b is losing its primordial atmosphere, we also expect observable Lyman $\alpha$ transits at 1215.6 \AA\:with HST \citep{gj436}. To estimate this, we assumed that the escaping hydrogen from the atmosphere of the planet would fill up it's Hill sphere, a lower limit on the geometry of the escaping atmosphere \citep{lymanalpha}. The Hill sphere is defined by the Hill radius \citep{Tremaine_2023}:
\begin{equation}
    R_{H}=a\left(\frac{M_{p}}{3M_{s}}\right)^{1/3}
\end{equation}
From this, we found $R_{H,6054b}=57.8$ $R_{\oplus}$. If we assume that the Lyman $\alpha$ transit is due to the entire Hill sphere of escaping hydrogen, we estimate a Lyman $\alpha$ transit depth of 10.2\% for TOI-6054b.

To estimate the HST Lyman $\alpha$ transit depth uncertainty, we rescaled the error on the depth of GJ 436 b from \cite{gj436}, assuming it is due to pure photon noise once again. We used the GALEX NUV magnitude in the 2000 \AA\:to 3000 \AA\:range as a proxy, since GJ 436 was measured in this catalog \citep{galex}. However, TOI-6054 is not in this catalog, so we used measurements TIC 70227585 instead, for the reasons described in Section \ref{sec:he}, and rescale it for distance. From this analysis, we estimate a Lyman $\alpha$ transit depth error of roughly 0.1\% for the TOI-6054 system. This is two orders of magnitude lower than our estimated transit depth for TOI-6054b, so we conclude that Lyman $\alpha$ transit observations with HST of TOI-6054b are feasible.

\section{Conclusions} \label{sec:conc}

We present the discovery and characterization of a pair of sub-Neptunes orbiting TOI-6054 using a combination of photometric and spectroscopic measurements as a part of the ORCAS survey. We found that TOI-6054b has a period of $7.501437 \pm 0.000061$ days, a radius of $2.65 \pm 0.15$ $R_{\oplus}$, a mass of $12.4 \pm 1.7$ $M_{\oplus}$, and an equilibrium temperature of $1360 \pm 33$ K, and that TOI-6054c has a period of $12.56364 \pm 0.00013$ days, a radius of $2.81 \pm 0.18$ $R_{\oplus}$, a mass of $9.2 \pm 2.0$ $M_{\oplus}$, and an equilibrium temperature of $1144 \pm 28$ K. The host star is about to evolve onto the sub-giant branch, and consequently, we expect that TOI-6054b could be actively losing its atmosphere. Additionally, we find that the pair of planets is very likely in a 5:3 mean motion resonance. This system is interesting from both a formation and dynamical standpoint and warrants future follow-up observations. The system is scheduled to be observed in TESS Sector 86, and the new data will likely help better constrain both the transit parameters and the strength of the dynamical interactions between the planets. We also plan to continue RV observations of the system to investigate the long-term background trend in the data.

\section*{Acknowledgements}

This paper contains data taken with the NEID instrument, which was funded by the NASA-NSF Exoplanet Observational Research (NN-EXPLORE) partnership and built by Pennsylvania State University. NEID is installed on the WIYN telescope, which is operated by the National Optical Astronomy Observatory, and the NEID archive is operated by the NASA Exoplanet Science Institute at the California Institute of Technology. NN-EXPLORE is managed by the Jet Propulsion Laboratory, California Institute of Technology under contract with the National Aeronautics and Space Administration.

This research has made use of the Exoplanet Follow-up Observation Program (ExoFOP; DOI: 10.26134/ExoFOP5; \citealt{exofop}) website and the NASA Exoplanet Archive, which are both operated by the California Institute of Technology, under contract with the National Aeronautics and Space Administration under the Exoplanet Exploration Program.

I.J.M.C. acknowledges support from NASA ROSES grant 24-XRP24\_2-0084, and from NASA Keck grant 80NSSC24K1096.

J.R.L. acknowledges financial support from a Fluno Graduate Fellowship through the University of Wisconsin--Madison.

D.R.C. acknowledges partial support from NASA Grant 18-2XRP18\_2-0007. This research has made use of the Exoplanet Follow-up Observation Program (ExoFOP; DOI: 10.26134/ExoFOP5) website, which is operated by the California Institute of Technology, under contract with the National Aeronautics and Space Administration under the Exoplanet Exploration Program.

J.J.L. acknowledges support from NASA ROSES grant 24-XRP24\_2-0020.

B.S.S. and M.V.G. acknowledge the support of M.V. Lomonosov Moscow State University Program of Development.

T.D. acknowledges support from the McDonnell Center for the Space Sciences at Washington University in St. Louis.

R.L. acknowledges support from NASA through the NASA Hubble Fellowship grant \#HST-HF2-51559.001-A awarded by the Space Telescope Science Institute, which is operated by the Association of Universities for Research in Astronomy, Inc., for NASA, under contract NAS5-26555.

J.M.A.M. is supported by the National Science Foundation (NSF) Graduate Research Fellowship Program (GRFP) under Grant No. DGE-1842400. J.M.A.M. and N.M.B. acknowledge support from NASA’S Interdisciplinary Consortia for Astrobiology Research (NNH19ZDA001N-ICAR) under award number 19-ICAR19\_2-0041.

Some of the observations in this paper made use of the High-Resolution Imaging instrument ‘Alopeke and were obtained under Gemini LLP Proposal Number: GN/S-2021A-LP-105. ‘Alopeke was funded by the NASA Exoplanet Exploration Program and built at the NASA Ames Research Center by Steve B. Howell, Nic Scott, Elliott P. Horch, and Emmett Quigley. Alopeke was mounted on the Gemini North telescope of the international Gemini Observatory, a program of NSF’s OIR Lab, which is managed by the Association of Universities for Research in Astronomy (AURA) under a cooperative agreement with the National Science Foundation. on behalf of the Gemini partnership: the National Science Foundation (United States), National Research Council (Canada), Agencia Nacional de Investigación y Desarrollo (Chile), Ministerio de Ciencia, Tecnología e Innovación (Argentina), Ministério da Ciência, Tecnologia, Inovações e Comunicações (Brazil), and Korea Astronomy and Space Science Institute (Republic of Korea).

\section*{Data Availability}

The TESS observations which were used in this paper are publicly available on the MAST archive: \dataset[10.17909/jt54-t744]{http://dx.doi.org/10.17909/jt54-t744}. The high resolution images are available on ExoFOP: \dataset[10.26134/ExoFOP5]{https://doi.org/10.26134/ExoFOP5}. The processed RV data from NEID underlying this article are reported in Table \ref{tab:rv}.

\appendix
\section{Posteriors from the Joint Fit to the Data}

\begin{figure}[!b]
    \centering
    \includegraphics[width=0.9\linewidth]{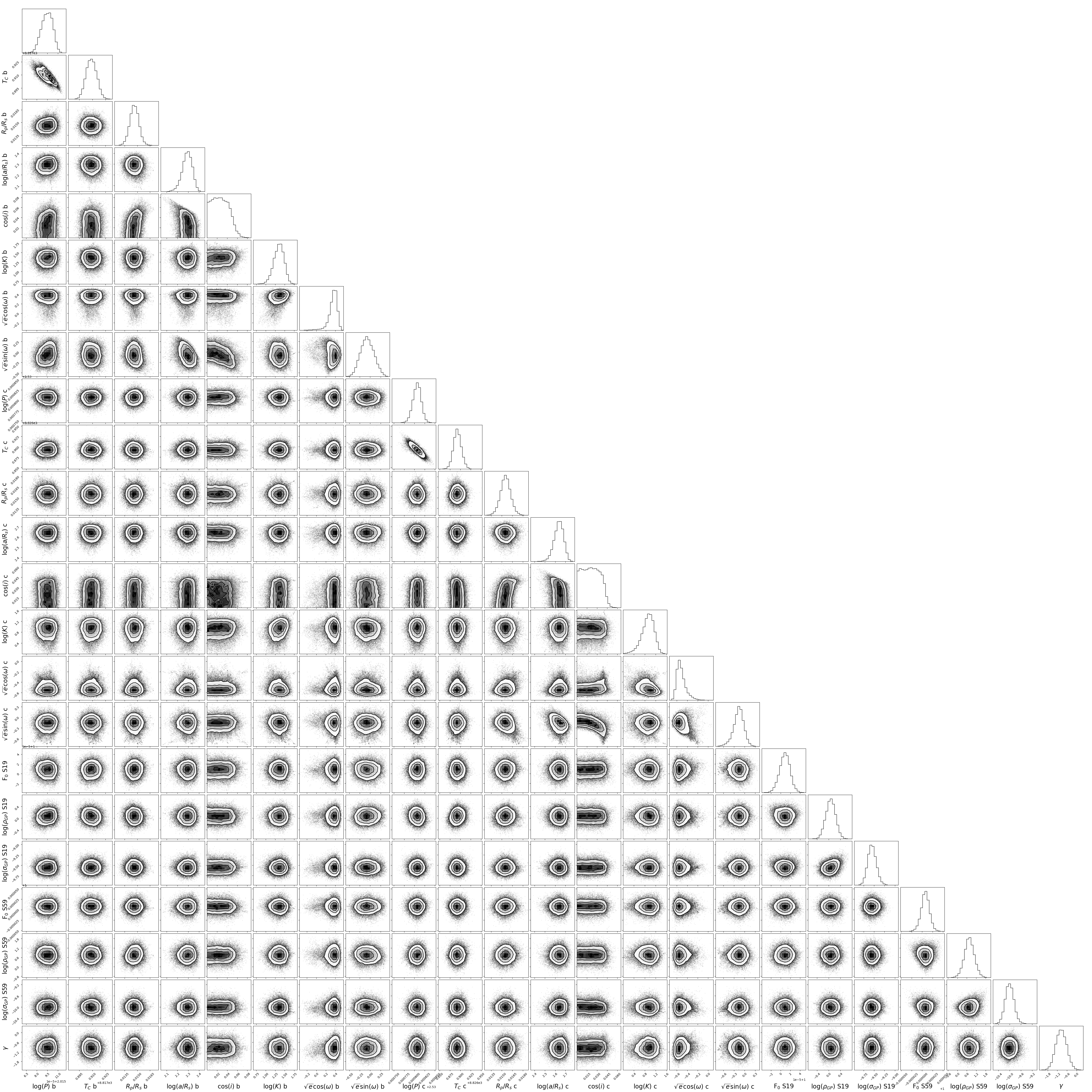}
    \caption{Corner plot of the posterior distributions of all fitted parameters in our joint fit described in Section \ref{sec:orbpar}. Parameters labeled with a ``b" correspond to TOI-6054b, and those with a ``c" correspond to TOI-6054c. Light curve fitting parameters specific to Sectors 19 and 59 are labeled with ``S19" and ``S59", respectively. The median values and 1$\sigma$ uncertainties for these parameters are reported in Table \ref{tab:plpar}.}
    \label{fig:corner}
\end{figure}

\newpage

\bibliography{bib}{}
\bibliographystyle{aasjournal}

\end{document}